\documentclass[fleqn,usenatbib]{mnras}

\usepackage{newtxtext,newtxmath}

\usepackage[T1]{fontenc}
\usepackage{hyperref}
\DeclareRobustCommand{\VAN}[3]{#2}
\let\VANthebibliography\thebibliography
\def\thebibliography{\DeclareRobustCommand{\VAN}[3]{##3}\VANthebibliography}

\usepackage{graphicx}	
\usepackage{amsmath}	
\usepackage{longtable}
\usepackage{booktabs}

\title[Discovery of Pulsating Components]{Discovery of $\delta$~Sct Components in Eclipsing Binary Systems IQ CMa, AW Men and W Vol}

\author[B. Ula\c{s} and C. Ulusoy]{
Burak Ula\c{s}$^{1}$\thanks{E-mail: burak.ulas@comu.edu.tr},
Ceren Ulusoy$^{2}$
\\
$^{1}$Department of Space Sciences and Technologies, Faculty of Sciences, \c{C}anakkale Onsekiz Mart University, Terzio\v{g}lu Campus, \c{C}anakkale, TR-17100,Turkey\\
$^{2}$Department of Physics and Astronomy, Botswana International University of Science and Technology (BIUST), Plot~10017, Private Bag 16, Palapye, Botswana
}

\date{Accepted XXX. Received YYY; in original form ZZZ}

\pubyear{0000}

\begin{document}
\label{firstpage}
\pagerange{\pageref{firstpage}--\pageref{lastpage}}
\maketitle

\begin{abstract}
We present the first evidence on the $\delta$~Sct type pulsations of the primary components of three eclipsing binaries IQ CMa, AW Men and W Vol in the TESS field. A comprehensive investigation of the binary properties is conducted. The light curves of the systems are analysed and the frequency analyses are performed to residual data. The systems are compared to the binaries of the same morphological types, and the primaries are examined in contrast to the $\delta$~Sct type pulsators. The results show that the systems are oscillating eclipsing Algol-type systems. 
\end{abstract}




\section{Introduction}

Scientific research on Eclipsing Binary Systems (EBs) constitutes the backbone of astrophysics in many respects. First and foremost, EBs play a vital role in studying the fundamental stellar parameters and distances that can be determined more precisely than the data obtained for single stars.

EBs with pulsating components are the most advantageous targets for direct measurement of component masses and radii in the case of high-mass short period binaries \citep[e.g.][]{rib00, tor10}. Thus, these targets are critical to uncovering the physical phenomena of massive pulsating stars in EBs \citep{lam22,tka20,joh21,sou22,sim17}.

Pulsating components can be found in a broad range of spectral types and in the various group of instability strips ($\delta$~Sct and $\beta$~ Cep stars) across the HR diagram \citep{lia17}. Over the last decade, our understanding of binary stars has remarkably improved since a large number of EBs with pulsating stars were discovered in the $\delta$~Sct Scuti regime by the {{\it Kepler}}/K2 \citep{bor10,gil10,how14} and TESS \citep{ric15} missions.

$\delta$~Sct stars are Population I stars situated on the main-sequence at the extension of the Cepheid instability strip with A0-F5 spectral types and effective temperatures of 6700–8900K \citep{sam18,cha13}. Their masses are between 1.5 and 2.5 M$\sun$, which correspond to the main sequence stage of core hydrogen or shell
burning. The oscillations of this type of pulsators are driven by {\it $\kappa$- mechanism} in radial and non-radial pressure and gravity modes or both \citep{gri10,uyt11}. The periods vary between about 0\fd 008-0\fd 42. \citep{cat15}. 

The study of EBs with $\delta$~Sct components can provide accurate asteroseismic models which are linked to the short-term stellar evolution dynamic because they undergo a series of several physical phenomena such as
tidal interactions between the components, mass transfer stages and pulsations \citep{sam18,lam21,lam22}. Particularly, in close binary systems, tidally-excited non-radial pulsations can also be involved in the stellar evolution process \citep{aer21}.

The literature has very limited information about the systems and all targets are mainly published in several catalogue studies such as Gaia DR2, DR3 \citep{gai18,bra21}, the binary star catalogue of \citet{avv13} and the TESS eclipsing binary catalogue of \cite{mor21}. \citet{wri03} indicated an effective temperature value, 7580~K, for IQ~CMa based on the spectral type, A8V. \citet{sha18} listed temperature and $\log g$ (7250~K and 3.8) values for W~Vol based on high resolution spectroscopy. The radial velocity limit for the target was catalogued by \citet{rei20} and the mid-infrared flux and diameter were given by \citet{cru19}. The lack of detailed studies, although the light curves show evident oscillation-like effects, motivated us to investigate the systems. This is the first report that presents the first evidence on pulsation characteristics of primary components, as well as the first comprehensive binary modeling in the literature.

\section{Light Curve Data}\label{lcdata}
The TESS light curves of three eclipsing binary systems were collected through Mikulski Archive for Space Telescopes (MAST) portal\footnote{\url{https://mast.stsci.edu/portal/Mashup/Clients/Mast/Portal.html}}. The data from sectors 6 and 33 for IQ~CMa, 2, 4 and 5 for AW~Men, and 2, 5 and 6 for W~Vol were used. Data files cover datapoints of 32308, 52304 and 51324 for IQ~CMa, AW~Men and W~Vol, respectively, with no NULL data. Since the Per-search Data Conditioning Simple Aperture Photometry (PDC SAP), which are used for detecting planetary transits, may contain some unphysical distortions, the Simple Aperture Photometry (SAP) fluxes were used during the analyses. SAP fluxes ($F_i$) from the data files were extracted and converted to magnitude values ($m_i$) using the formula $m{_i}=-2.5\log F{_i}$ and the TESS magnitudes \citep{sta19} which are 8$\overset{m}{.}$964, 11$\overset{m}{.}$862 and 10$\overset{m}{.}$272 for IQ~CMa, AW~Men and W~Vol, respectively. A linear trend was extracted from the light curves, when necessary, in order to avoid artificial effects.

\begin{figure*}
\centering
\includegraphics[width=\textwidth]{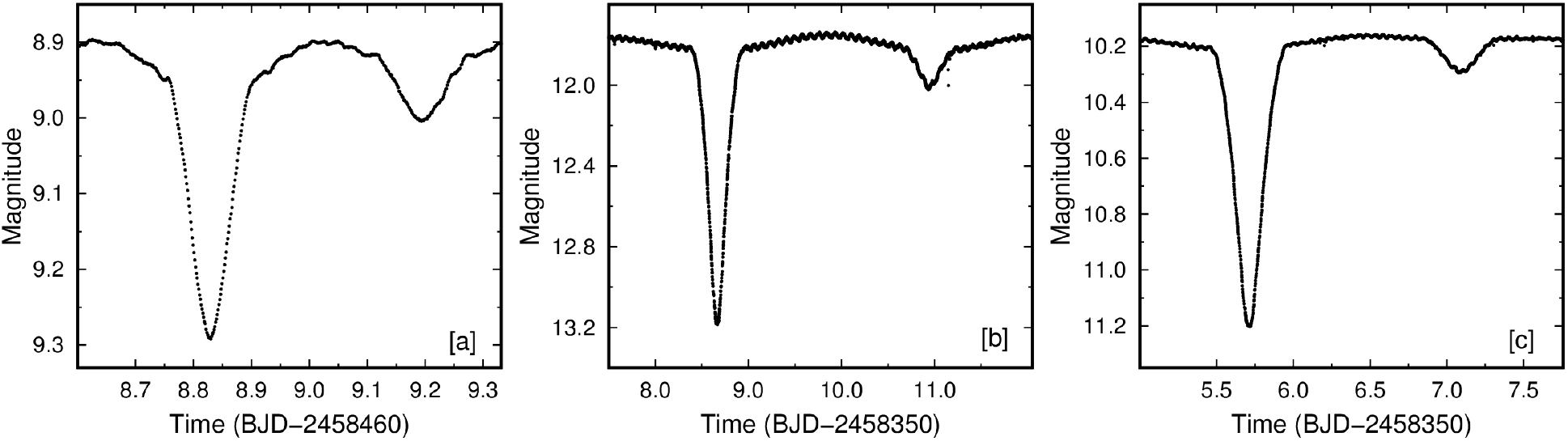}
\caption{TESS light curve in one orbital period for IQ~CMa (a), AW~Men (b) and W~Vol (c).}
\label{maglc}
\end{figure*}

The light curves of the systems resemble a typical eclipsing binary light curve affected by the waveform of the oscillations, mainly observed in the maximum phases. The dominance of oscillations during secondary minima has also led us to consider pulsating components to be primary. The depths of the primary minima are 0$\overset{m}{.}$33 for IQ~CMa, 1$\overset{m}{.}$37 for AW~Men and 0$\overset{m}{.}$97 for W~Vol. The primary minima last 3\fh 2, 11\fh 9 and 11\fh 3, while the duration of the secondary minima are 3\fh 4, 12\fh 5 and 10\fh 8 for IQ~CMa, AW~Men and W~Vol, respectively. Fig.~\ref{maglc} shows the light curves covering only one orbital period of the binaries with the aim of displaying the oscillating effect clearly.

\section{Modelling Binaries}\label{binsec}

\begin{figure}
\centering
\includegraphics[scale=0.75]{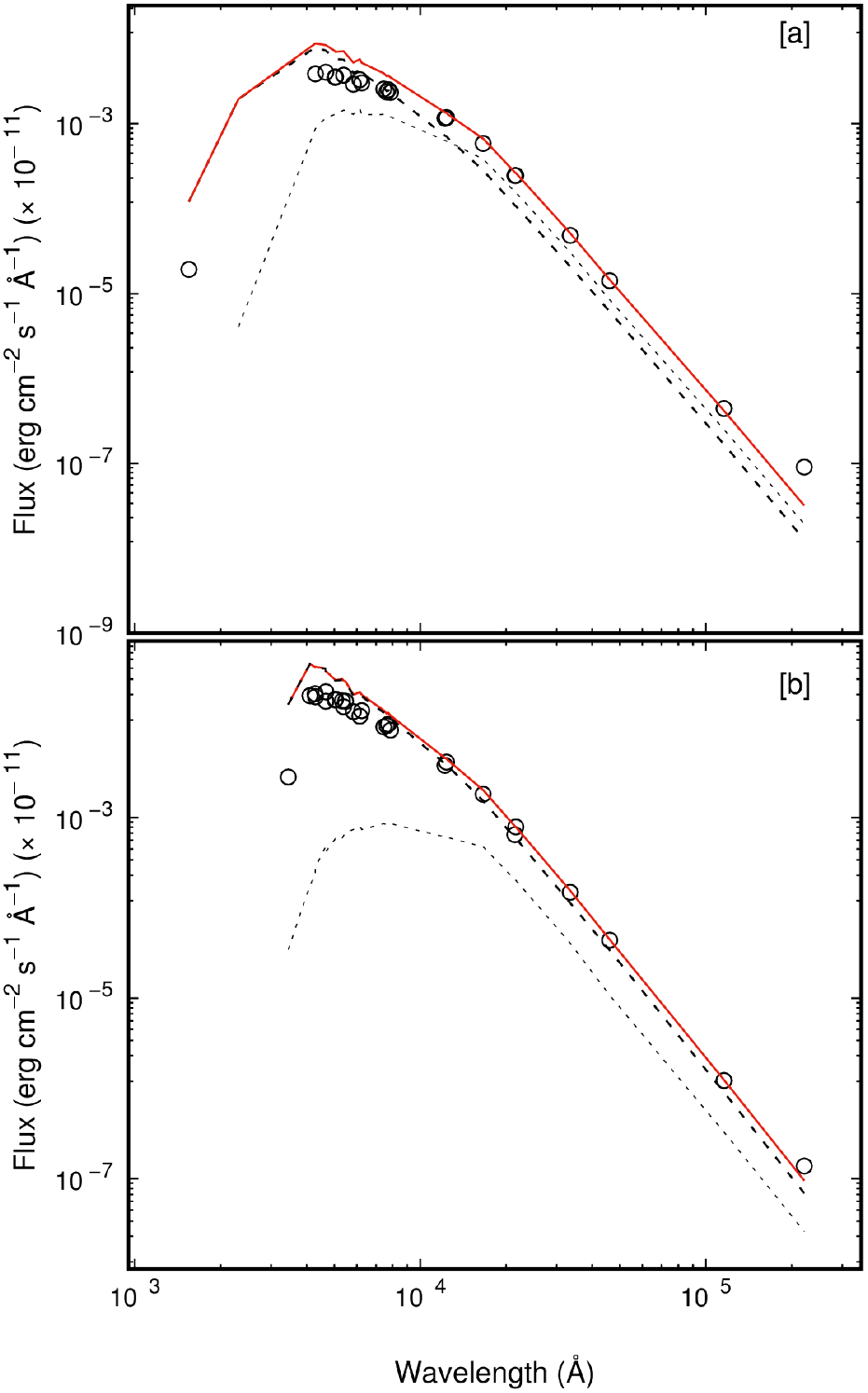}
\caption{Observational points and the fit from the SED analyses for AW~Men (a) and W~Vol (b). The dashed and dotted lines represent the best model fits for the primary and the secondary components, respectively. The total model fits are indicated by the red lines.}
\label{sed}
\end{figure}

In order to derive the final parameters of the observed binary system, the effective temperature of the primary component is one of the critical parameters in starting a light curve analysis. Since our targets have not been studied individually in the literature, there is no sufficient information on the temperatures of AW~Men and W~Vol based on their spectral types. Therefore, their temperature determinations were performed by constructing the spectral energy distributions (SEDs) by using the Virtual Observatory SED Analyzer\footnote{\url{http://svo2.cab.inta-csic.es/theory/vosa/index.php}} \citep[VOSA,][]{bay08} based on photometric data of the VizieR database \citep{och00} as similar to previously carried out by \citet{ula22}. The tool requires a series of intervals for initial effective temperature, $\log~g$ and metallicity values to achieve Vgf$_b$, a value for estimating the goodness of fit, given as \citep{bay08}:
\begin{equation}
{\text{Vgf}}_{b}=\frac{1}{N-n_p}\sum_{i=1}^{n}\frac{(Y_{i,o}-M_{d}Y_{i,m})^{2}}{b_{i,o}^{2}}
\end{equation}
where $N$ is the number of photometric data points, $n_p$ is the number of fitted parameters, $Y_o$ is the observed flux, $Y_m$ is the predicted flux and $M_d$ is the multiplicative dilution factor. The $b_o$ parameter satisfies $b_o = 0.1 Y_o$ if the error in the observational flux is smaller than $0.1 Y_o$, elsewise, it is equal to the observational flux. Therefore, Vgf$_b$ is defined as modified reduced $\chi^2$ calculated by forcing the observed flux error to be larger than 10\% of the observed flux. Its values smaller than 10–15 correspond to a good fit. We selected the binary fit option (fitting data based on two components) with a parameter-grid search to obtain the optimized Kurucz atmosphere model \citep{kur79}. The SED diagrams are given in Fig.~\ref{sed}. The initial and resulting values are also discussed in the corresponding subsections.

\begin{figure}
\centering
\includegraphics[width=0.4\textwidth]{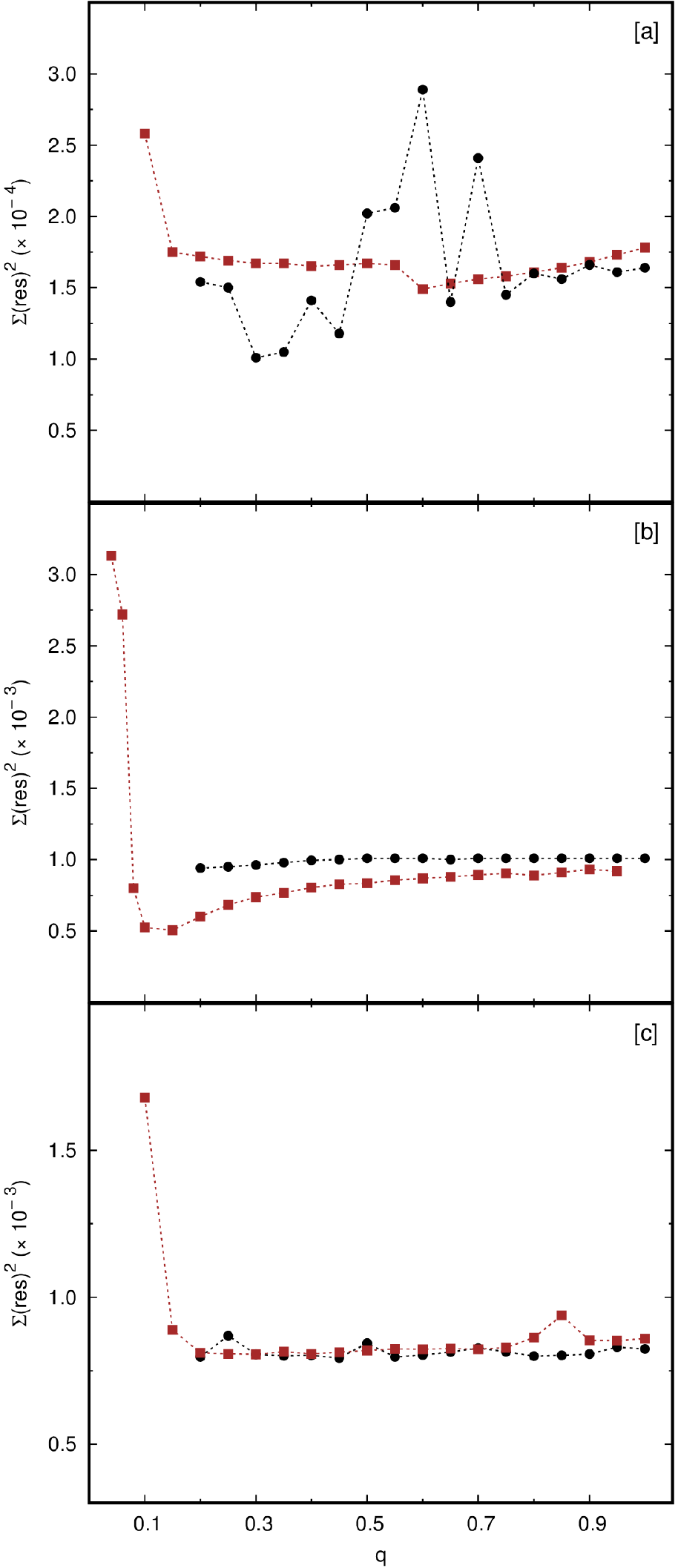}
\caption{Results of the $q$-search with detached (circles) and semidetached (squares) assumptions for IQ~CMa (a), AW~Men (b) and W~Vol (c).}
\label{qsearch}
\end{figure}

Since there is no comprehensive research available on the determination of the binary parameters in the literature,  it is essential to establish an initial photometric mass ratio value for each target to achieve physically significant results by means of light curve analyses. Thus, $q$-search processes were applied to binned light curves covering 1000 data points using a 2015 version of the Wilson-Devinney code \citep{wil71,wil20}. The q-searches were carried out in accordance with two morphological assumptions, detached and semidetached, and the results are presented in Fig.\ref{qsearch}. The squared residuals indicate that IQ~CMa is a detached binary, while two other targets are semidetached systems. However, the nearly constant structures in Fig.~\ref{qsearch} show that the light curve is almost insensitive to the mass ratio in detached configurations. Moreover, it is slightly dependent on the mass ratio in semidetached binaries, especially for systems with partial eclipses, as stated by \cite{ter05}. Therefore, additional system characteristics should also be considered as well as the results of the light curve solutions in two configurations should be compared to determine morphologies as discussed in the following subsections.

The orbital periods of the systems were derived by averaging the consecutive times of minima of the same type. Since two of our systems, AW~Men and W~Vol, are in the TESS Southern Continuous Viewing Zone (SCVZ) and were observed during years 1 and 3, we used light curves of 1800 s and 600 s cadenced Full Frame Images (FFIs) during the determination of the period in order to derive more accurate values. Frequency analyses were also carried out on the light curves to validate the value derived for the orbital period. The times of minima calculated in this study were conducted to a period analysis by assuming that the variation is linear, namely $O - C = \Delta T_{0} + \Delta P \times E$, where $\Delta P$ and $\Delta T_0$ are the variations in orbital period, the time of minimum light and $E$ is the cycle number. The calculated period values are listed in Table~\ref{lc_tab} with the times of minimum light which were adopted during our light curve solutions. Standard errors ($SE$) in orbital periods were derived by using the formula \mbox{\( \displaystyle SE = \sigma / \sqrt{N} \)}, where $\sigma$ is the standard deviation and $N$ is the number of period values determined from the consecutive minima.

The light curves were analysed using the {\tt PHOEBE} software \citep{pri05} which bases the Wilson-Devinney method \citep{wil71} to obtain the stellar parameters from the input data. The light curves of AW~Men and W~Vol contain 52304 and 51324 data points, respectively. However, we restricted the number of input data to 50000, since it is the maximum number of data that can be entered into the program. During the analyses, the albedos ($A_{1}$, $A_{2}$) were derived from  \citet{ruc69} and gravity darkening coefficients ($g_{1}$, $g_{2}$) were adopted from \citet{zei24} and \citet{luc67} by bearing in mind that the granulation boundary for main sequence stars located at about F0 spectral type \citep{gra89}. Logarithmic limb darkening coefficients ($x_{1}$ and $x_{2}$) were taken from the catalogue by \citet{cla17} based on the initial temperatures and the gravities of the components.

\subsection{IQ~CMa}

For IQ~CMa, an effective temperature value, 7580~K, was given by \citet{wri03} based on the spectral type A8V remarked by \citet{hou88}. In our solution, 7580~K was adopted as the temperature of the primary component. The value derived from the orbital period was 0\fd 73138 using the aforementioned methodology. Although the minimal $q$-search supports the detached configuration, the shape of the light curve resembles a typical Algol binary and calls into question the detached configuration. In addition, our analysis shows that the system contains the K type secondary companion, which is generally seen in characteristic Algol-type binary systems. We, therefore, analysed the light curve in two different configurations, detached and semidetached, in order to clarify the morphology.

\begin{figure}
\centering
\includegraphics[scale=0.85]{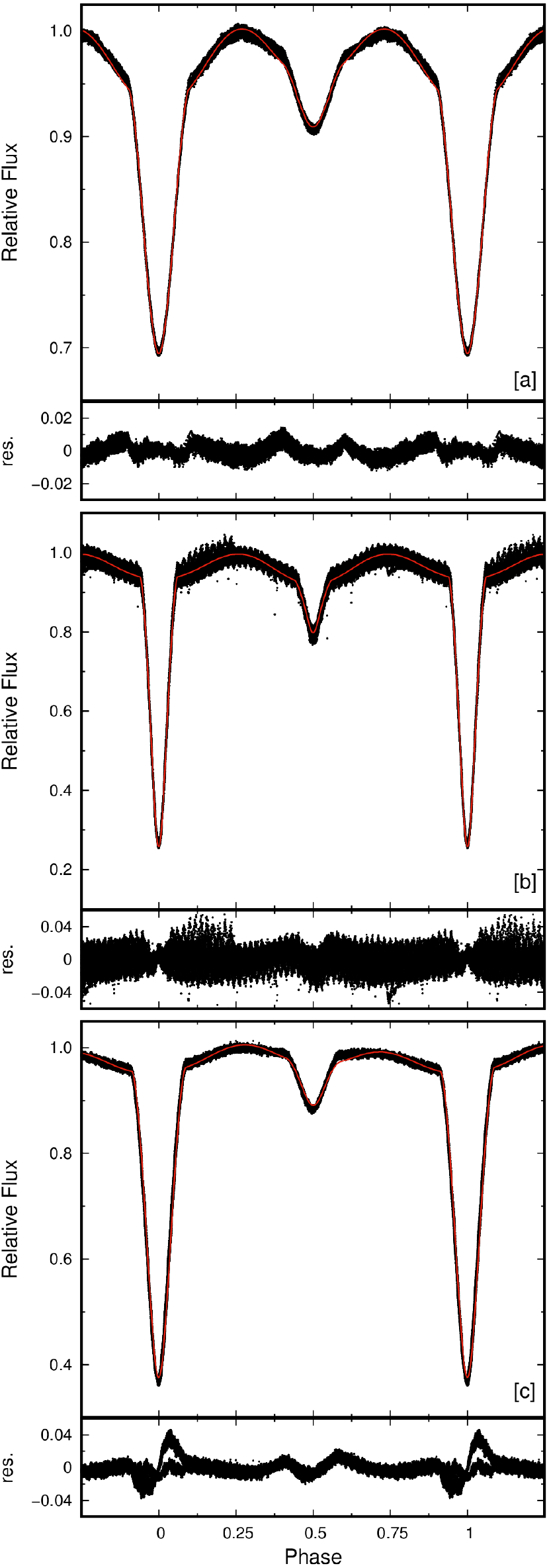}
\caption{Observational TESS and synthetic light curves of IQ~CMa (a), AW~Men (b) and W~Vol (c) plotted with the residuals.}
\label{lcs}
\end{figure}

The initial mass ratio values were selected to be 0.6 and 0.3 for the analyses with semidetached and detached assumptions, respectively, which are in accord with our $q$-search. We adjusted the inclination $i$, the temperature of the secondary component $T_2$, mass ratio $q$, the surface potential values of the primary components $\Omega_1$ and luminosity of the primary component $L_1$  during the semidetached solutions. $\Omega_2$ of secondary was also set as an adjustable parameter in the analysis with the detached approach. The albedo of the secondary component was assigned as a free parameter for a couple of runs to achieve a better fit during the solutions. Finally, we produced more reasonable and accurate results in the semidetached configuration, while the detached approximation induces larger squared residuals and thus a remarkably poor fit. Considering all of the arguments noted above, we refer to the system as a semidetached binary. The observations are compared with the synthetic light curve in Fig.~\ref{lcs} and the result of the analysis is presented in Table~\ref{lc_tab}.

\begin{table}
\centering
\caption{Results of the light curve analyses. $r_1$ and $r_2$ denote the fractional radii of the components. The numbers in parentheses refer to the standard deviations, 3$\sigma$ for the last digits of light parameters, while they indicate the standard errors for T$_0$ and $P$. Uncertainties in $T_1$ values are given as obtained in the SED analyses.}
\label{lc_tab}
\begin{tabular}{lccc}
\hline
Parameter & IQ~CMa & AW~Men & W~Vol \\
\hline           
$i$ ${({^\circ})}$      & 68.56(1)   	& 86.47(3) 		& 80.78(2) 		\\       
$q$                     & 0.555(1) 		& 0.111(1) 		& 0.333(1)		\\         
$T_1$ (K)               & 7850(125) 	& 7500(125) 	& 7500(125) 	\\   
$T_2$ (K)               & 5016(45)		& 5342(43)		& 4780(58)		\\           
$\Omega _{1}$           & 3.831(3) 		& 6.02(1)		& 4.339(9) 		\\         
$\Omega _{2}=\Omega_{cr}$  & 2.980 	& 1.965         & 2.538		\\  
$\frac{L_1}{L_1 +L_2}$  & 0.873(1) 		& 0.753(1)		& 0.871(2)		\\  
$r_1$                   & 0.310(2) 		& 0.169(5)		& 0.251(3) 		\\    
$r_2$                   & 0.327(1) 		& 0.211(2)		& 0.286(1)		\\ 
$x_{1},~x_{2}$ 		    & 0.522, 0.658 		& 0.531, 0.670 		& 0.573, 0.715		\\
T$_0$ (BJD) 		    & 1468.82927(1)~ 	& 1358.6655(8)~ 	& 1328.1299(8)~ 	\\    
$P$~(days)   		    & 0.73138(2)  		& 4.5524(2) 		& 2.7581(1)  	\\        
\hline
\end{tabular}
\end{table}

\subsection{AW~Men}

The effective temperature of AW~Men was determined by the SED analysis, as indicated previously. During the analysis, the initial temperature and $\log g$ range for both the primary and secondary components were taken as 3500-10000~K and 2.5-5.0, respectively. The distance of the system was adopted as 1108$\pm$25~pc following the value based on Gaia~DR2 \citep{gai16,gai18} parallax. The extinction, A$_{\nu}=$0.0287$^m$, was derived by using the Galactic Dust Reddening and Extinction interface of NASA/IPAC Infrared Science Archive\footnote{\url{https://irsa.ipac.caltech.edu/applications/DUST/}}. We assumed that the stars are in solar abundance. The best fit values were $T_1 = 7500\pm125~K$, $\log g_1 = 3.5\pm0.25$, $T_2 = 4750\pm125~K$ and $\log g_1 = 3.0\pm0.25$. The Vgf$_b$ of the analysis was found to be 0.214, which corresponds to a good fit, as explained earlier. The orbital period of the system was calculated as 4\fd 5524 using the method described above. Therefore, we set these parameters as the initial values for the light curve solution. 

Considering the small difference between the minimum squared residuals of q-searches, the relatively large orbital period of the system and the weak dependence of the light curve to the mass ratio, we applied the light curve solutions with two different configurational assumptions, detached and semidetached. The free parameters were the same as in the IQ~CMa solution. The detached model results in a notably poor fit, especially around the edges of the secondary minimum. The $\chi^2$ of the detached solution was significantly greater than the semidetached solution. Since the best output of binary properties is essential for the frequency analysis (Sec.~\ref{freqsec}), we agreed to proceed with the semidetached configuration. The initial mass ratio was taken as 0.15, the value from the $q$-search. The observed and computed light curves are plotted with the residuals in Fig.~\ref{lcs} and the resultant light parameters are presented in Table~\ref{lc_tab}. The system is a relatively long period Algol with a low mass ratio like S~Equ \citep{soy07}, T~LMi \citep{oka77} and RV~Oph \citep{wal70} which were previously reported by the researchers as well as listed in the catalogue of Algol-type binaries by \citet{bud04}.

\subsection{W~Vol}

The SED analysis for the system was done by setting the temperature and $\log g$ domains to 3500-10000~K and 2.5-5.0, respectively, for both of the components. The distance parameter was 697(10)~pc, the value corresponding Gaia~DR2 \citep{gai16,gai18} parallax. A$_{\nu}$ value in the direction of the system was fixed at 0.4785$^m$. The calculation was made by the solar abundance approach. The results indicate that $T_1 = 7500\pm125~K$, $\log g_1 = 3.0\pm0.25$, $T_2 = 4000\pm125~K$ and $\log g_1 = 2.5\pm0.25$ with the Vgf$_b$ value of 1.14, coinciding with the criterion of good fit. The temperature can be considered as close to the value given by \citet{sha18}, 7290~K. The orbital period of the system was calculated as 2\fd 7581. 

Our $q$-search supports detached configuration. However, the very small difference in the squared residuals and the insensitivity of the light curve to the mass ratio in detached systems prompted us to solve the light curve in both detached and semidetached approximation. The free parameters were identical to the ones used in the analyses of other systems. Our detached solution corresponded that the secondary component is oversized according to its mass and does not follow the mass-radius relation as expected from the components of detached binaries. This situation led us to conclude that stellar parameters are unphysical considering that the components of detached binaries evolve independently of each other, and it is thus expected that they strictly follow the mass-radius relation. Therefore, we decided to conduct a further analysis with a semidetached approximation. During the semidetached solution, a hot spot on the primary component due to mass transfer was hypothesized to fit the difference between maximum phases of the light curve more accurately and achieve a better extraction of the binary model. Hot spot approximation was previously introduced in the literature to model the similar asymmetry in the light curves of Algol-type systems such as TW~Dra \citep{wal78} and WZ~Crv \citep{vir11}, AV~Hya and DZ~Cas \citep{yan12}. Furthermore, \cite{rod04} modeled the light curve of Algol-type binary RZ~Cas by including a hot spot on the $\delta$-Sct type pulsating primary component. \cite{pet12} remarked the presence of a hot spot on the pulsating component of WX~Dra. The primary component of KIC~10063044, a $\gamma$~Dor type pulsator, was also assumed to host a hot spot during the light curve analysis made by \cite{ozd20}. In our solution, the parameters of the hot spot on the surface of the primary was co-latitude $\beta$=80$^{\circ}$, longitude $\lambda$=280$^\circ$, fractional radius $r$=25$^\circ$ and the temperature factor $t$=1.1. We concluded that the semidetached configuration is more plausible morphology for the system based on our effort in analyzing the light curve. The observed and synthetic light curves are shown with the residuals in Fig.~\ref{lcs} while the binary properties are listed in Table~\ref{lc_tab}.

\section{Frequency Analyses}\label{freqsec}

\subsection{IQ~CMa}

\begin{figure*}
\centering
\includegraphics[width=\textwidth]{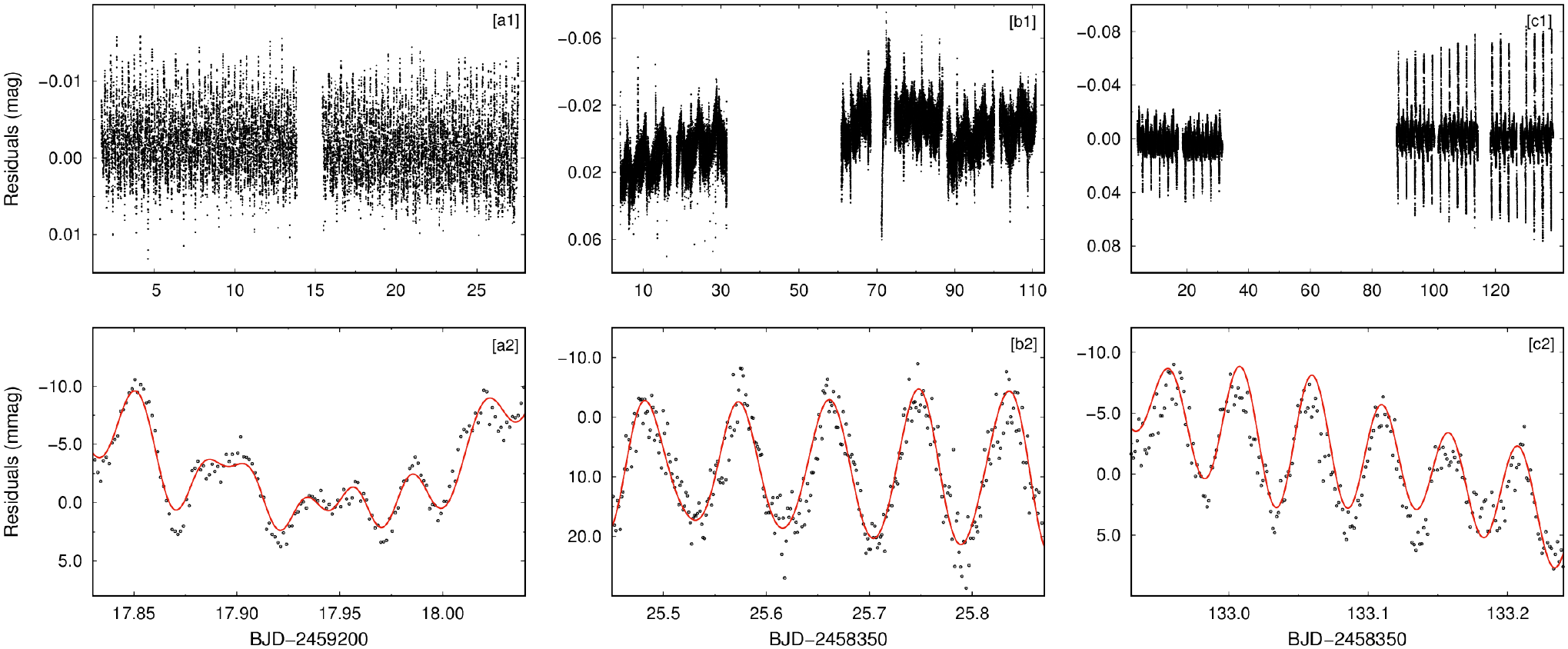}
\caption{The residual light curves (upper panels) and the final fits (lower panels, red lines) obtained from the frequency analysis for IQ~CMa (a1, a2), AW~Men (b1, b2) and W~Vol (c1, c2).}
\label{res}
\end{figure*}

\begin{figure}
\centering
\includegraphics[scale=0.75]{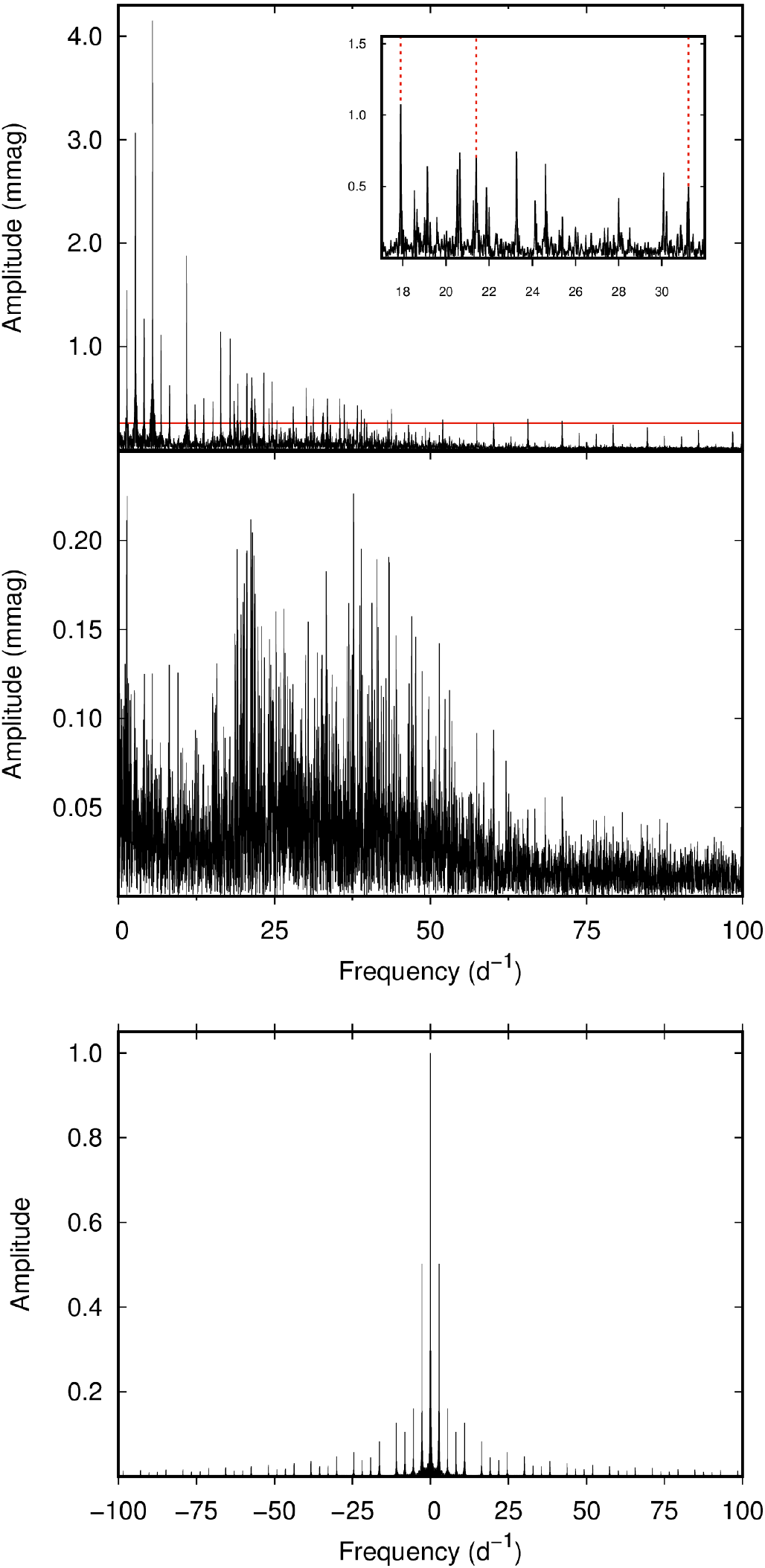}
\caption{Amplitude spectra in the beginning of the frequency analysis and after prewhitened all calculated frequencies are represented for IQ~CMa in the top and the middle panels. The last panel is the spectral window of the data. The horizontal red line refers to the significance level.}
\label{f3760}
\end{figure}

After subtracting the binary model from the observational data, we applied frequency analysis to the maximum phases of the residuals using the {\tt PERIOD04} software \citep{len05}. The program is based on Fourier analysis and calculates amplitude spectra for a given frequency interval. We only used the data from sector 33 in the frequency analysis to avoid the effect of a significant time gap between the two available sectors. As a result, a total of 10500 data points were imported into the program. The frequency range was set to 0-100~d$^{-1}$. The analysis shows that the residuals can be represented by three genuine and 26 combination frequencies with the signal-to-noise ratio (SNR) is higher than the critical limit of the software, 4.0. The genuine frequencies are located between 17 and 32~d$^{-1}$ and listed in Table~\ref{gentab}. All genuine frequencies are higher than 5~d$^{-1}$ where the frequencies of $\delta$~Sct type pulsators are observed \citep{gri10}. The pulsation constants, $Q$, for the genuine frequencies were calculated as 0.0400, 0.0335, and 0.0229 for $f_5$, $f_8$ and $f_{10}$, respectively, based on the absolute parameters (Table~\ref{tababs})  by using the equation where $\frac{\rho}{\rho_{\odot}}$ is the mean density of the pulsating primary in solar unit and $P$ is the period of pulsation: 

\begin{align}
    Q=P\sqrt{\left( \frac{\rho}{\rho_{\odot}}\right)}.
\end{align}
The pulsation constants for $f_8$ and $f_{10}$ are inside the range of the interval for $\delta$~Sct stars \citep[i.e. 0.015~$<Q<~$0.035;][]{bre00} although it is close to the boundary of the criteria for $f_5$. In addition, the effective temperature of the primary, 7850~K, is between A-F spectral types, the spectral range covers $\delta$~Sct stars. Combining the above arguments with the semidetached geometry from the light curve solution (see Sec.~\ref{binsec}), it can be considered that the primary component of the system may be considered to be a $\delta$~Sct-type pulsator, and the binary is an Oscillating Eclipsing Systems of Algol-type (oEA) as defined by \citet{mkr02}. The residuals are plotted and the agreement of the fit from frequency analysis is shown in Fig.~\ref{res}. Fig.~\ref{f3760} represents the amplitude spectra and the spectral window.

\begin{table}
\centering
\caption{Genuine frequencies calculated during the frequency analyses. Parameters $f$, $A$, $\phi$, and SNR stand for frequency, amplitude, phase, and signal-to-noise ratio, respectively. The least-square uncertainties are given in the last digits. See the Appendix for the list of possible combination frequencies.}
\label{gentab}
\begin{tabular}{lcccc}
\hline
& $f$ (d$^{-1}$) & $A$ (mmag) & $\phi$  &  SNR \\
\hline
IQ~CMa &  & &   &   \\
\cmidrule{1-1}
$f_{5} $&17.9013(3)&0.00109(2)&0.326(2)&26.8\\
$f_{8} $&21.3914(5)&0.00074(2)&0.092(4)&11.3\\
$f_{10} $&31.2378(8)&0.00047(2)&0.764(6)&9.9\\
\hline
AW~Men &  & &   &   \\
\cmidrule{1-1}
$f_{1}$&0.00657(1)&0.01190(3)&0.3810(4)&82.4\\
$f_{2}$&11.57049(2)&0.00842(3)&0.6935(6)&114.1\\
\hline
W~Vol &  & &   &   \\
\cmidrule{1-1}
$f_{2}$&0.00597(4)&0.00172(2)&0.100(2)&22.5\\
$f_{3}$&19.38688(3)&0.00239(2)&0.861(1)&30.7\\
$f_{10}$&16.29856(9)&0.00074(2)&0.797(4)&10.9\\
$f_{11}$&20.40910(11)&0.00064(2)&0.044(4)&8.1\\
$f_{13}$&15.35579(12)&0.00056(2)&0.042(5)&9.1\\
$f_{14}$&15.83035(12)&0.00056(2)&0.165(5)&8.6\\
\hline
\end{tabular}
\end{table}

\subsection{AW~Men}
\begin{figure}
\centering
\includegraphics[scale=0.75]{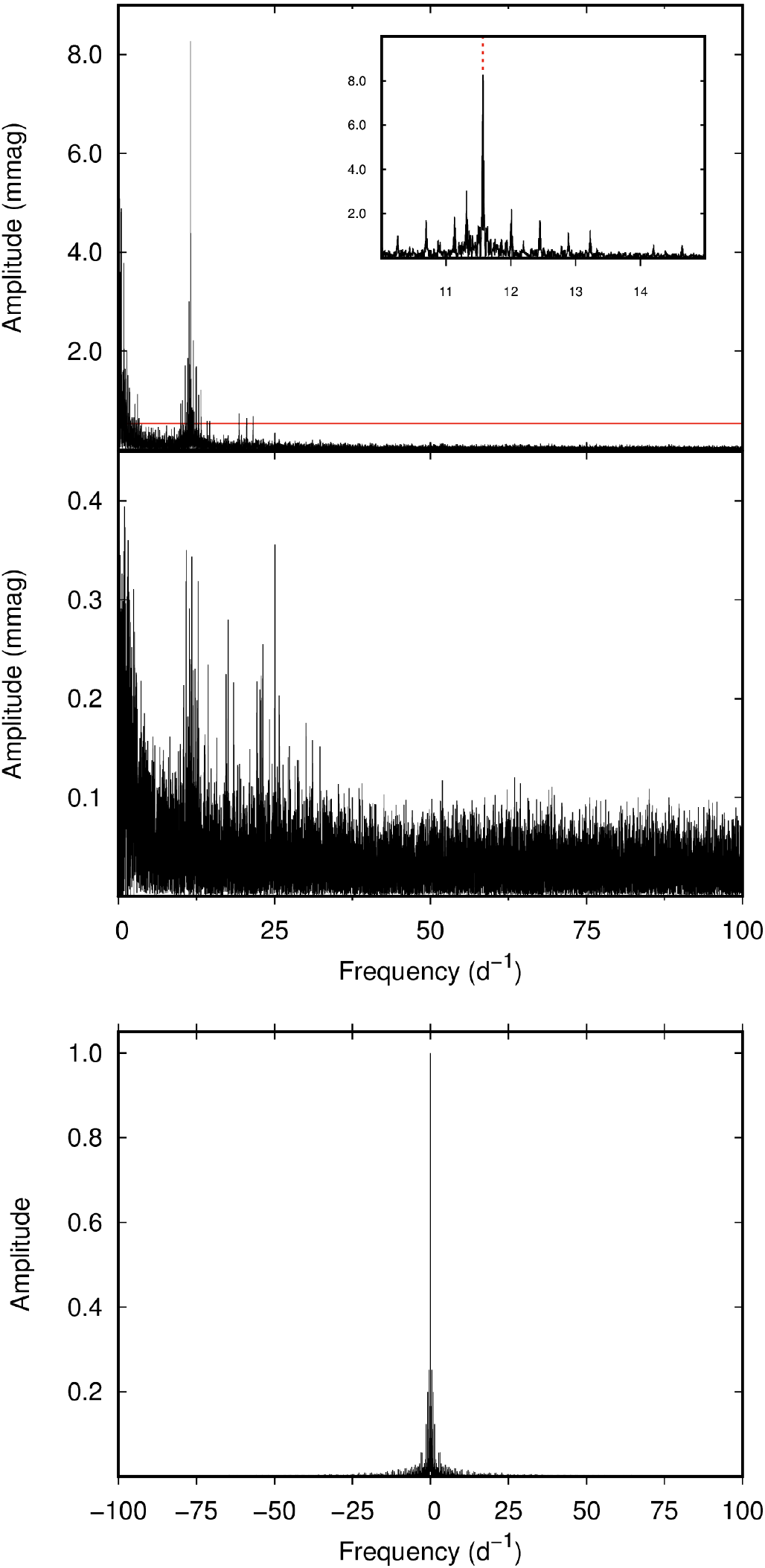}
\caption{Same as Fig.~\ref{f3760}, but for AW~Men.}
\label{f3919}
\end{figure}

The frequency analysis was employed on 38520 data points at maximum phases, yielded after the extraction of the binary model. The analysis, which covers the frequency range between 0 and 100~d$^{-1}$ corresponds to two genuine and 39 combination frequencies with SNR higher than 4.0 (Table~\ref{gentab}). The genuine frequency $f_1$, 0.00657~d$^{-1}$, is likely the result of the instrumental effects rising from different pixel sensitives and the variation of the aperture size by sectors. The other frequency, $f_2$, satisfied the range of $\delta$~Sct regime proposed by \citet{bre00}. The pulsation constant, $Q$, was calculated as 0.0306 for $f_2$ using the absolute parameters of the primary and Eq.~2. This value ensures that the frequency complies with the criteria for $\delta$~Sct type pulsators, as it was mentioned above. Moreover, following the SED analysis, the effective temperature of the primary, 7500~K, coincides with the stars having the spectral types of A-F. The evidences presented thus far support the idea that AW~Men is a semidetached binary having a $\delta$~Sct type primary component. In other words, the system is found to be an oEA system. The residuals of the binary extraction and the final fit from the analysis are shown in Fig.~\ref{res} while the amplitude spectra is plotted in Fig.~\ref{f3919}.

\subsection{W~Vol}
\begin{figure}
\centering
\includegraphics[scale=0.75]{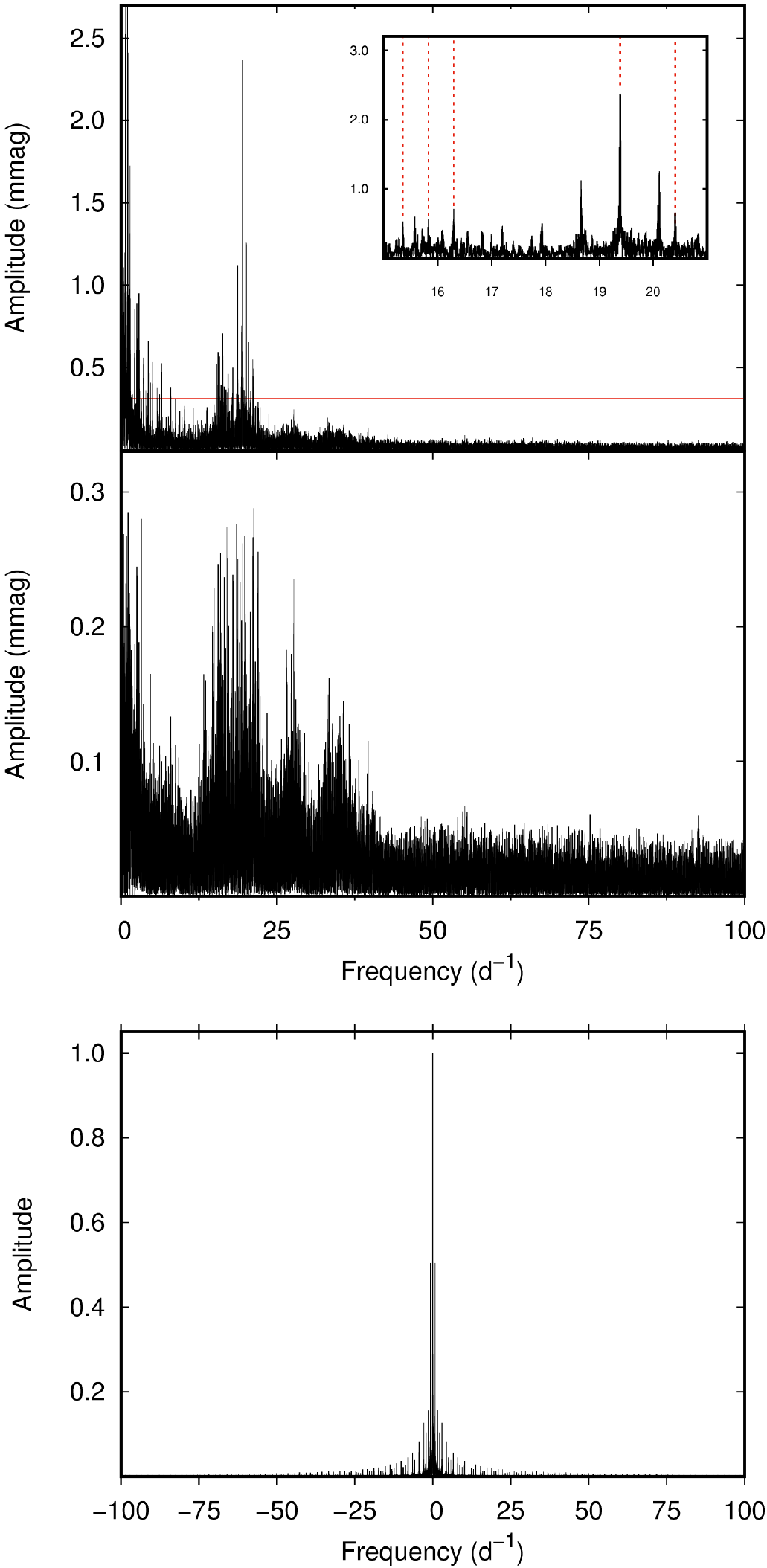}
\caption{Same as Fig.~\ref{f3760}, but for W~Vol.}
\label{f3006}
\end{figure}

Six genuine and 29 combination frequencies were the result of the frequency analysis which is applied to 29816 data points from the residual light curve of the system. The analysis was employed in the 0 and 100~d$^{-1}$ range. The maximum value of the genuine frequencies is about 20~d$^{-1}$. $f_2$ of the genuine frequencies is probably rising from the instrumental effects, while the rest are higher than 5~d$^{-1}$ and also between 5 and 80~d$^{-1}$, the $\delta$~Sct criteria of both \citet{gri10} and \citet{bre00}. Additionally, the pulsation constants (Eq.~2) are 0.121, 0.0144 and 0.0115 for $f_3$, $f_{10}$ and $f_{11}$, respectively, which are very close to the lower edge of the $\delta$~Sct interval \citep[i.e. 0.015~$<Q<~$0.035;][]{bre00}, while the constants for $f_{13}$ and $f_{14}$ (0.0152 and 0.0148) can be considered within the range. In addition, the SED analysis corresponds to the effective temperature of the primary is, 7500~K, indicating an A-F type star. As the case clearly demonstrates that the primary component of the system can be considered as a $\delta$~Sct type pulsator. The residuals light curve and the fit from the frequency analysis are plotted in Fig.~\ref{res}. Fig.~\ref{f3006} represents the amplitude spectra.

\section{Conclusion}

We present the first evidence on the oscillations of primary components of the systems in question. This is also the first study that presents detailed analyses of the light curves of the selected targets. The absolute parameters based on the light curve solutions were calculated by using the {\tt{AbsParEB}} program \citet{lia15a} and given in Table~\ref{tababs}. It is important to bear in mind that the Wilson-Devinney method underestimates the uncertainties, thus the standard errors in Table~\ref{lc_tab} and \ref{tababs} are unphysical. The masses for the primary components of the systems were estimated based on their effective temperature and the $\log g$ values among 204930 binary star models created by using the Binary Star Evolution code \citep[{\tt{BSE}},][]{hur02,hur13} with the initial values of eccentricity between 0 and 0.5, the orbital period between 0.5~d and 7~d, the mass of the primary component, $M_1$, between 0.5 M$_{\odot}$ and 5.0 M$_{\odot}$, the mass of the secondary component between 0.1 M$_{\odot}$ and $M_1$, and with solar abundance assumption. 

\begin{figure*}
\centering
\includegraphics[scale=0.45]{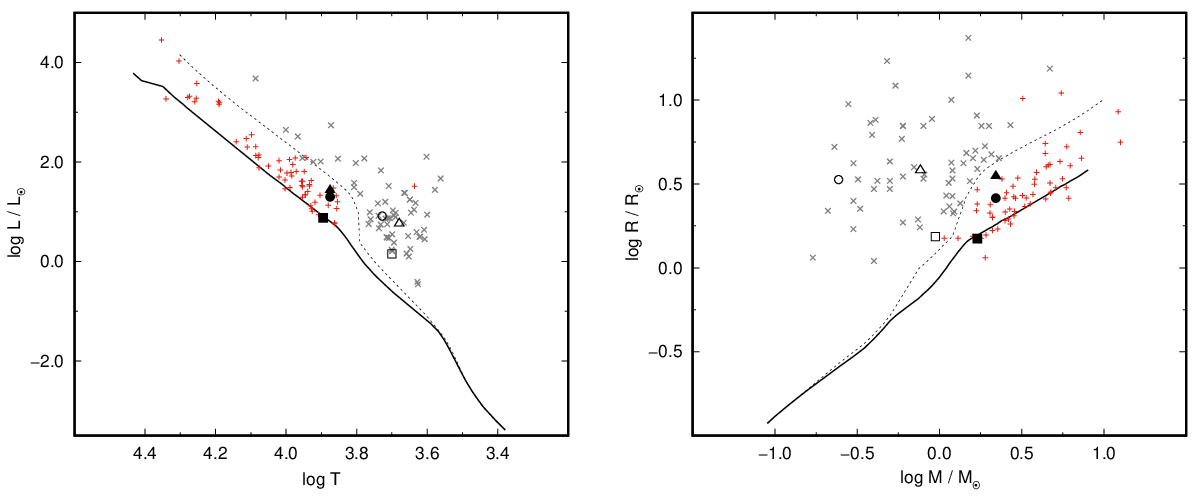}
\caption{Location of the components on the Hertzsprung-Russell diagram (left) and the mass-radius plane (right). Filled and open signs denote the primary and secondary components, while the square, circle and triangle refer to IQ~CMa, AW~Men and W~Vol. Plus and crosses (grey and tan) present the primary and secondary components of known Algols \citep{iba06}. The data for ZAMS (thick solid line) and TAMS (dashed line) are taken from \citet{bre12} with $Y=0.267$ and $Z=0.01$ abundance.}
\label{fighrmr}
\end{figure*}

Based on the resulting light parameters of IQ~CMa (Table \ref{lc_tab}) the filling factor, $f=r/r_L$, for the primary component is found to be about 94\% by using the equation for the volume radius of the Roche Lobe \citep{egg83}:

\begin{align}
    r_{L}=\frac{0.49q^{2/3}}{0.6q^{2/3}+ln(1+q^{1/3})}.
\end{align}
This outcome led us to the possibility that the system is a near contact binary. According to the classification of \citet{sha94}, the Roche lobe filling secondary indicate that the system is a member of FO Virginis class, the progenitors of A-type W~UMa systems, although the lack of significant difference between two maxima of the light curve and the secondary component does not seem to be sufficiently over-sized in the Hertzsprung-Russell diagram and the mass-radius plane (Fig.~\ref{fighrmr}).  

\begin{figure*}
\centering
\includegraphics[scale=0.75]{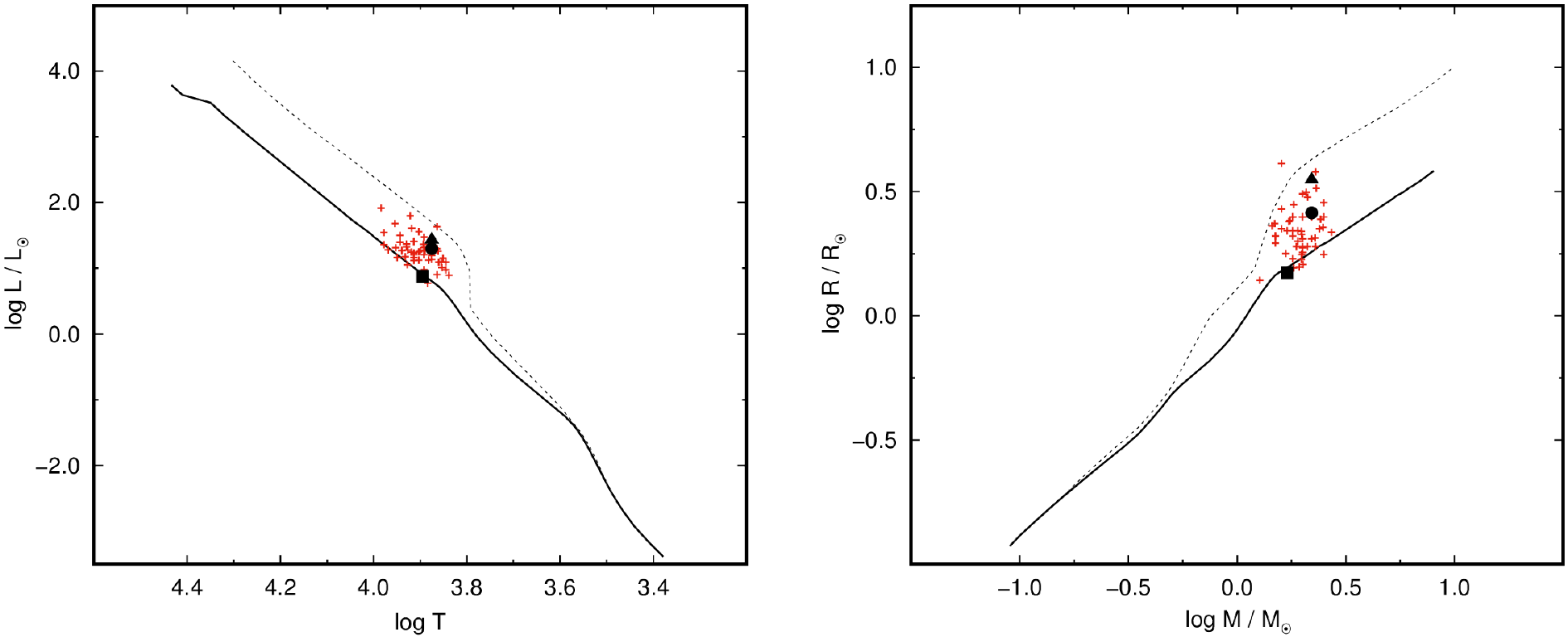}
\caption{Same as Fig.~\ref{fighrmr}, but for primary components of our targets and known $\delta$~Sct components (red plus signs) in semidetached eclipsing binaries given by \citet{lia17}.}
\label{fighrmrdsct}
\end{figure*}

\begin{figure*}
\centering
\includegraphics[width=\textwidth]{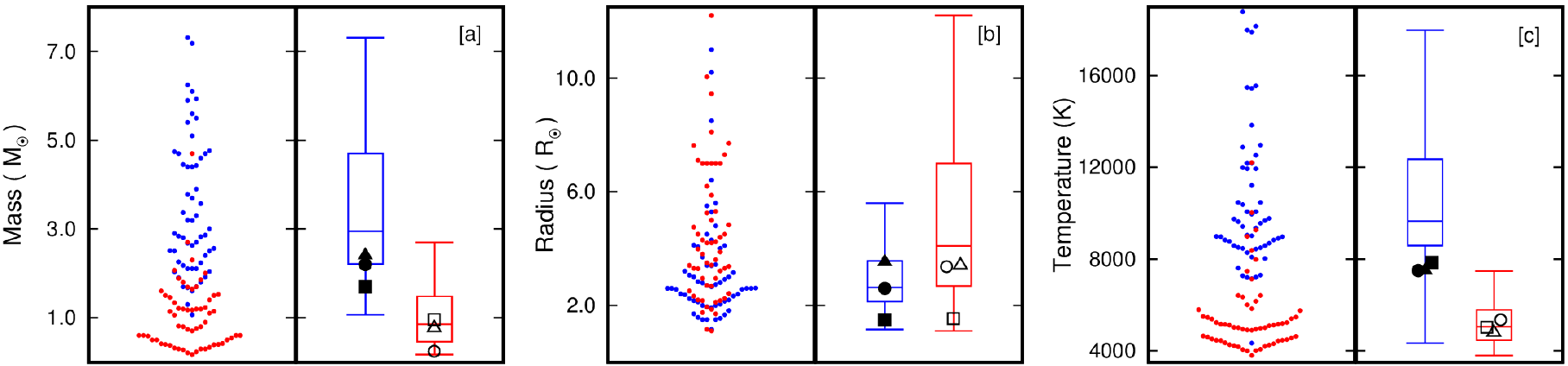}
\caption{Distributions of mass, radius and temperature values of the primary and secondary components of Algols are shown on the left halves of a, b, and c based on the data given by \citet{iba06}. The right panels correspond to the box plots whose properties are given in Table~\ref{tabbox}. Blue and red circles refer to primary and secondary components, respectively. The other symbols are the same as those displayed in Fig.~\ref{fighrmr}. Some extreme values are not included in the plot areas for the sake of visibility, although boxes are calculated using the whole data.}
\label{figbox}
\end{figure*}

The targets were compared to well known binaries of the same type on the Hertzsprung-Russell diagram and the mass-radius plane in Fig.~\ref{fighrmr}. The systems show a good agreement with the other Algol-type binaries. Fig.~\ref{fighrmrdsct} illustrates the comparison of the pulsating components to known $\delta$~Sct type components in semidetached binary systems \citep{lia17} on the Hertzsprung-Russell diagram and the mass-radius plane. The locations of the primaries are in accord with the other $\delta$~Sct pulsators. Additionally, for statistical comparison, the position of our targets are shown in the mass, radius and temperature distributions for Algol-type systems in Fig.~\ref{figbox}. The box plots \citep{krz14} were drawn and lower quartile ($Q1$), median ($Q2$) and upper quartile ($Q3$) values were calculated. The interquartile range, the height of the boxes in the figure, were derived using the relation $IQR=Q3-Q1$. The results show that the parameters of the systems are inside the distributions, although deviations from the $IQR$ are observed. The summary of box plots is given in Table~\ref{tabbox}. 

\begin{table}
\centering
\caption{Absolute parameters of the systems. The standard errors are given in parentheses for the last digits.}\label{tababs}
\setlength{\tabcolsep}{1pt}
\begin{tabular}{lccc}
\hline
Parameter & ~~IQ~CMa~~ & ~~AW~Men~~ & ~~W~Vol\\
\hline
M$_1$ (M$_{\sun}$) &1.7        & 2.2       & 2.3\\
M$_2$ (M$_{\sun}$) &0.944(2)   & 0.244(2)  & 0.766(2)\\
R$_1$ (R$_{\sun}$) &1.49(2)    & 2.6(3)    &3.52(2)\\
R$_2$ (R$_{\sun}$) &1.53(5)    & 3.36(9)   &3.82(2)\\
L$_1$ (L$_{\sun}$) &7.6(2)     & 20(4)     &27(1)\\
L$_2$ (L$_{\sun}$) &1.42(1)    & 8.2(4)    &5.79(8)\\
$a $ (R$_{\sun}$) & 4.834(3)   & 15.94(2)  &12.306(1)\\
\hline                                      
\end{tabular}
\end{table}

\begin{table}
\centering
\caption{Summary of the box plots of 62 Algol-type binary systems whose data were provided by \citet{iba06}. $Q1$, $Q2$, $Q3$ and $IQR$ refers to lower quartile, median, upper quartile and interquartile range, respectively. Masses and radii are given in solar unit, while the temperatures are in K.}
\label{tabbox}
\begin{tabular}{lcccccc}
\hline
 & $Q1$ & $Q2$ & $Q3$ & min. & max. & $IQR$\\
\hline
M$_1$ &2.24   &  3.00 &  4.70 &  1.07 & 7.31  &2.46 \\
M$_2$ &0.48    & 0.81 & 1.46 & 0.17  &2.70   &0.98 \\
R$_1$ &2.15     &2.65  &3.43  &1.15  &5.28  &1.28 \\
R$_2$ &2.70      &4.20   &7.00   &1.10   &12.20  &4.30 \\
T$_1$ &8590 & 9638 & 12190 & 4325 & 15560 & 3600 \\
T$_2$ &4457  &5023  &5754  &3647  &7465 & 1298\\
\hline
\end{tabular}
\end{table}

\begin{figure}
\centering
\includegraphics[width=\columnwidth]{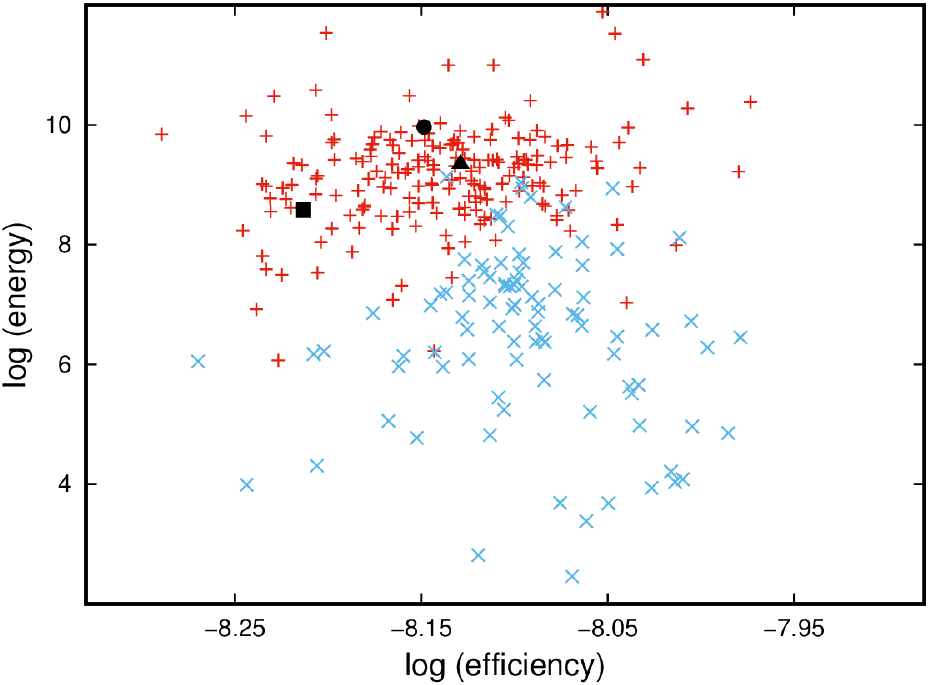}
\caption{Location of the primary components on $log$(efficiency)-$log$(energy) diagram. Plus (red) and cross (blue) signs refer to, $\delta$~Sct and $\gamma$~Dor type stars, respectively. The other symbols are the same as those displayed in Fig.~\ref{fighrmr}. The data are from \citet{uyt11}.}
\label{figee}
\end{figure}

The $\delta$~Sct type candidacy of the primaries was also inspected by locating the stars on the energy-efficiency diagram, purposed by \citet{uyt11}. The diagram, constructed by using {\it Kepler} targets and $\delta$~Sct and $\gamma$~Dor type pulsators, distinguishes between two distinct groups. Energy is defined as $energy\equiv (A_{max}\zeta_{max})^2$, where, $A_{max}$ is the highest amplitude mode (in ppm) of the frequency value of $\zeta_{max}$. The efficiency, on the other hand, is related to effective temperature and $\log g$ through $efficiency\equiv (T_{eff} ^{3} \log g)^{-2/3}$. Fig.~\ref{figee} shows the agreement of our targets with the $\delta$~Sct type pulsators as they gathered on the $\delta$~Sct region of the diagram.

\begin{figure}
\centering
\includegraphics[width=\columnwidth]{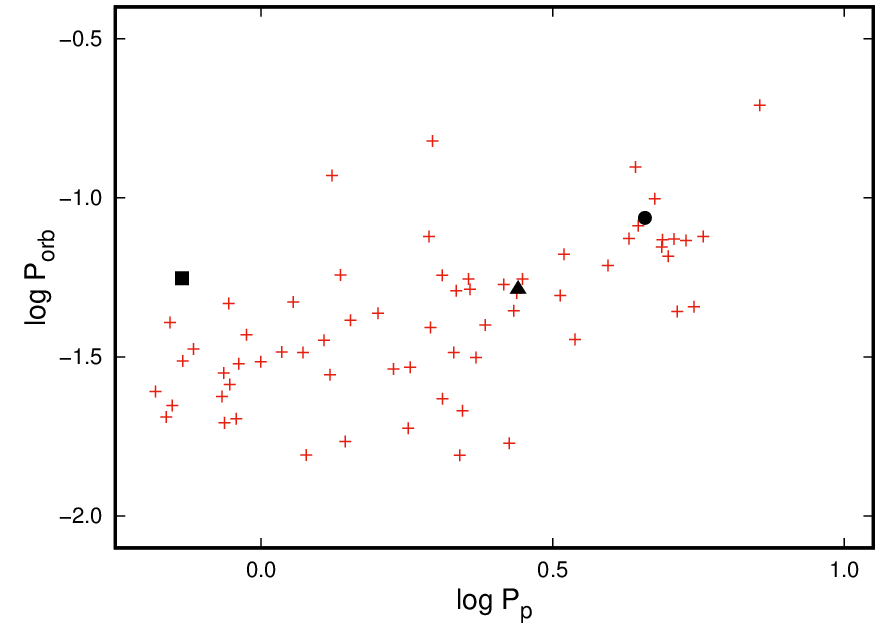}
\caption{Location of the systems on $\log P_{orb}-\log P_p$ diagram with the known semidetached systems having $P_{orb} < 13^{d}$. Plus (red) signs refer to the semidetached binaries. The other symbols are the same as those displayed in Fig.~\ref{fighrmr}. The data are given by \citet{lia17}}
\label{figpp}
\end{figure}

The relation between the orbital period and the fundamental frequency of the targets is examined in Fig.~\ref{figpp} by placing on $\log P_{orb}-\log P_p$ diagram with the semidetached systems having orbital periods smaller than 13~days and whose data are given by \citet{lia17}. The locations of the targets agree with the general trend and endorse that the primaries are $\delta$~Sct type pulsators. 

Considering the overall findings and results from the SED analyses, light curve solutions, frequency analyses and comparisons in the study, we report that the systems are eclipsing binaries with pulsating components. Furthermore, they were found to be oEA systems. In future investigations, the inclusion of spectroscopic observations is recommended to relieve the binary and oscillation properties more precisely.

\section*{Acknowledgements}

The authors would like to thank to the referee for the constructive comments and recommendations. The numerical calculations reported in this paper were partially performed at TUBITAK ULAKBIM, High Performance and Grid Computing Center (TRUBA resources). This paper includes data collected with the TESS mission, obtained from the MAST data archive at the Space Telescope Science Institute (STScI). Funding for the TESS mission is provided by the NASA Explorer Program. STScI is operated by the Association of Universities for Research in Astronomy, Inc., under NASA contract NAS 5–26555. This work has made use of data from the European Space Agency (ESA) mission
{\it Gaia} (\url{https://www.cosmos.esa.int/gaia}), processed by the {\it Gaia}
Data Processing and Analysis Consortium (DPAC,
\url{https://www.cosmos.esa.int/web/gaia/dpac/consortium}). Funding for the DPAC
has been provided by national institutions, in particular the institutions
participating in the {\it Gaia} Multilateral Agreement.This research has made use of the NASA/IPAC Infrared Science Archive, which is operated by the Jet Propulsion Laboratory, California Institute of Technology, under contract with the National Aeronautics and Space Administration. This publication makes use of VOSA, developed under the Spanish Virtual Observatory project supported by the Spanish MINECO through grant AyA2017-84089. VOSA has been partially updated by using funding from the European Union's Horizon 2020 Research and Innovation Programme, under Grant Agreement nº 776403 (EXOPLANETS-A). This research has made use of NASA's Astrophysics Data System. This research has made use of the VizieR catalogue access tool, CDS, Strasbourg, France.

\clearpage
\section*{Data Availability}

The data underlying this article are available at the MAST data archive at the Space Telescope Science Institute (STScI, \url{https://mast.stsci.edu/portal/Mashup/Clients/Mast/Portal.html})




\bibliographystyle{mnras}
\bibliography{Ulas_bib} 

\begin{thebibliography}{}
\makeatletter
\relax
\def\mn@urlcharsother{\let\do\@makeother \do\$\do\&\do\#\do\^\do\_\do\%\do\~}
\def\mn@doi{\begingroup\mn@urlcharsother \@ifnextchar [ {\mn@doi@}
  {\mn@doi@[]}}
\def\mn@doi@[#1]#2{\def\@tempa{#1}\ifx\@tempa\@empty \href
  {http://dx.doi.org/#2} {doi:#2}\else \href {http://dx.doi.org/#2} {#1}\fi
  \endgroup}
\def\mn@eprint#1#2{\mn@eprint@#1:#2::\@nil}
\def\mn@eprint@arXiv#1{\href {http://arxiv.org/abs/#1} {{\tt arXiv:#1}}}
\def\mn@eprint@dblp#1{\href {http://dblp.uni-trier.de/rec/bibtex/#1.xml}
  {dblp:#1}}
\def\mn@eprint@#1:#2:#3:#4\@nil{\def\@tempa {#1}\def\@tempb {#2}\def\@tempc
  {#3}\ifx \@tempc \@empty \let \@tempc \@tempb \let \@tempb \@tempa \fi \ifx
  \@tempb \@empty \def\@tempb {arXiv}\fi \@ifundefined
  {mn@eprint@\@tempb}{\@tempb:\@tempc}{\expandafter \expandafter \csname
  mn@eprint@\@tempb\endcsname \expandafter{\@tempc}}}

\bibitem[\protect\citeauthoryear{{Aerts}}{{Aerts}}{2021}]{aer21}
{Aerts} C.,  2021, \mn@doi [Reviews of Modern Physics]
  {10.1103/RevModPhys.93.015001}, \href
  {https://ui.adsabs.harvard.edu/abs/2021RvMP...93a5001A} {93, 015001}

\bibitem[\protect\citeauthoryear{{Avvakumova}, {Malkov}  \&
  {Kniazev}}{{Avvakumova} et~al.}{2013}]{avv13}
{Avvakumova} E.~A.,  {Malkov} O.~Y.,   {Kniazev} A.~Y.,  2013, \mn@doi
  [Astronomische Nachrichten] {10.1002/asna.201311942}, \href
  {https://ui.adsabs.harvard.edu/abs/2013AN....334..860A} {334, 860}

\bibitem[\protect\citeauthoryear{{Bayo}, {Rodrigo}, {Barrado Y Navascu{\'e}s},
  {Solano}, {Guti{\'e}rrez}, {Morales-Calder{\'o}n}  \& {Allard}}{{Bayo}
  et~al.}{2008}]{bay08}
{Bayo} A.,  {Rodrigo} C.,  {Barrado Y Navascu{\'e}s} D.,  {Solano} E.,
  {Guti{\'e}rrez} R.,  {Morales-Calder{\'o}n} M.,   {Allard} F.,  2008, \mn@doi
  [\aap] {10.1051/0004-6361:200810395}, \href
  {https://ui.adsabs.harvard.edu/abs/2008A&A...492..277B} {492, 277}

\bibitem[\protect\citeauthoryear{{Borucki} et~al.,}{{Borucki}
  et~al.}{2010}]{bor10}
{Borucki} W.~J.,  et~al., 2010, \mn@doi [Science] {10.1126/science.1185402},
  \href {https://ui.adsabs.harvard.edu/abs/2010Sci...327..977B} {327, 977}

\bibitem[\protect\citeauthoryear{{Brandt}}{{Brandt}}{2021}]{bra21}
{Brandt} T.~D.,  2021, \mn@doi [\apjs] {10.3847/1538-4365/abf93c}, \href
  {https://ui.adsabs.harvard.edu/abs/2021ApJS..254...42B} {254, 42}

\bibitem[\protect\citeauthoryear{{Breger}}{{Breger}}{2000}]{bre00}
{Breger} M.,  2000, \mn@doi [Baltic Astronomy] {10.1515/astro-2000-0122}, \href
  {https://ui.adsabs.harvard.edu/abs/2000BaltA...9..149B} {9, 149}

\bibitem[\protect\citeauthoryear{{Bressan}, {Marigo}, {Girardi}, {Salasnich},
  {Dal Cero}, {Rubele}  \& {Nanni}}{{Bressan} et~al.}{2012}]{bre12}
{Bressan} A.,  {Marigo} P.,  {Girardi} L.,  {Salasnich} B.,  {Dal Cero} C.,
  {Rubele} S.,   {Nanni} A.,  2012, \mn@doi [\mnras]
  {10.1111/j.1365-2966.2012.21948.x}, \href
  {https://ui.adsabs.harvard.edu/abs/2012MNRAS.427..127B} {427, 127}

\bibitem[\protect\citeauthoryear{{Budding}, {Erdem}, {{\c{C}}i{\c{c}}ek},
  {Bulut}, {Soydugan}, {Soydugan}, {Baki{\c{s}}}  \& {Demircan}}{{Budding}
  et~al.}{2004}]{bud04}
{Budding} E.,  {Erdem} A.,  {{\c{C}}i{\c{c}}ek} C.,  {Bulut} I.,  {Soydugan}
  F.,  {Soydugan} E.,  {Baki{\c{s}}} V.,   {Demircan} O.,  2004, \mn@doi [\aap]
  {10.1051/0004-6361:20034135}, \href
  {https://ui.adsabs.harvard.edu/abs/2004A&A...417..263B} {417, 263}

\bibitem[\protect\citeauthoryear{{Catelan} \& {Smith}}{{Catelan} \&
  {Smith}}{2015}]{cat15}
{Catelan} M.,  {Smith} H.~A.,  2015, {Pulsating Stars}.
Wiley VCH

\bibitem[\protect\citeauthoryear{{Chang}, {Protopapas}, {Kim}  \&
  {Byun}}{{Chang} et~al.}{2013}]{cha13}
{Chang} S.~W.,  {Protopapas} P.,  {Kim} D.~W.,   {Byun} Y.~I.,  2013, \mn@doi
  [\aj] {10.1088/0004-6256/145/5/132}, \href
  {https://ui.adsabs.harvard.edu/abs/2013AJ....145..132C} {145, 132}

\bibitem[\protect\citeauthoryear{{Claret}}{{Claret}}{2017}]{cla17}
{Claret} A.,  2017, \mn@doi [\aap] {10.1051/0004-6361/201629705}, \href
  {https://ui.adsabs.harvard.edu/abs/2017A&A...600A..30C} {600, A30}

\bibitem[\protect\citeauthoryear{{Cruzal{\`e}bes} et~al.,}{{Cruzal{\`e}bes}
  et~al.}{2019}]{cru19}
{Cruzal{\`e}bes} P.,  et~al., 2019, \mn@doi [\mnras] {10.1093/mnras/stz2803},
  \href {https://ui.adsabs.harvard.edu/abs/2019MNRAS.490.3158C} {490, 3158}

\bibitem[\protect\citeauthoryear{{Eggleton}}{{Eggleton}}{1983}]{egg83}
{Eggleton} P.~P.,  1983, \mn@doi [\apj] {10.1086/160960}, \href
  {https://ui.adsabs.harvard.edu/abs/1983ApJ...268..368E} {268, 368}

\bibitem[\protect\citeauthoryear{{Gaia Collaboration} et~al.,}{{Gaia
  Collaboration} et~al.}{2016}]{gai16}
{Gaia Collaboration} et~al., 2016, \mn@doi [\aap]
  {10.1051/0004-6361/201629272}, \href
  {https://ui.adsabs.harvard.edu/abs/2016A&A...595A...1G} {595, A1}

\bibitem[\protect\citeauthoryear{{Gaia Collaboration} et~al.,}{{Gaia
  Collaboration} et~al.}{2018}]{gai18}
{Gaia Collaboration} et~al., 2018, \mn@doi [\aap]
  {10.1051/0004-6361/201833051}, \href
  {https://ui.adsabs.harvard.edu/abs/2018A&A...616A...1G} {616, A1}

\bibitem[\protect\citeauthoryear{{Gilliland} et~al.,}{{Gilliland}
  et~al.}{2010}]{gil10}
{Gilliland} R.~L.,  et~al., 2010, \mn@doi [\pasp] {10.1086/650399}, \href
  {https://ui.adsabs.harvard.edu/abs/2010PASP..122..131G} {122, 131}

\bibitem[\protect\citeauthoryear{{Gray} \& {Nagel}}{{Gray} \&
  {Nagel}}{1989}]{gra89}
{Gray} D.~F.,  {Nagel} T.,  1989, \mn@doi [\apj] {10.1086/167505}, \href
  {https://ui.adsabs.harvard.edu/abs/1989ApJ...341..421G} {341, 421}

\bibitem[\protect\citeauthoryear{{Grigahc{\`e}ne} et~al.,}{{Grigahc{\`e}ne}
  et~al.}{2010}]{gri10}
{Grigahc{\`e}ne} A.,  et~al., 2010, \mn@doi [\apjl]
  {10.1088/2041-8205/713/2/L192}, \href
  {https://ui.adsabs.harvard.edu/abs/2010ApJ...713L.192G} {713, L192}

\bibitem[\protect\citeauthoryear{{Houk} \& {Smith-Moore}}{{Houk} \&
  {Smith-Moore}}{1988}]{hou88}
{Houk} N.,  {Smith-Moore} M.,  1988, {Michigan Catalogue of Two-dimensional
  Spectral Types for the HD Stars. Volume 4, Declinations -26{\textdegree}.0 to
  -12{\textdegree}.0.}.
University of Michigan

\bibitem[\protect\citeauthoryear{{Howell} et~al.,}{{Howell}
  et~al.}{2014}]{how14}
{Howell} S.~B.,  et~al., 2014, \mn@doi [\pasp] {10.1086/676406}, \href
  {https://ui.adsabs.harvard.edu/abs/2014PASP..126..398H} {126, 398}

\bibitem[\protect\citeauthoryear{{Hurley}, {Tout}  \& {Pols}}{{Hurley}
  et~al.}{2002}]{hur02}
{Hurley} J.~R.,  {Tout} C.~A.,   {Pols} O.~R.,  2002, \mn@doi [\mnras]
  {10.1046/j.1365-8711.2002.05038.x}, \href
  {https://ui.adsabs.harvard.edu/abs/2002MNRAS.329..897H} {329, 897}

\bibitem[\protect\citeauthoryear{{Hurley}, {Tout}  \& {Pols}}{{Hurley}
  et~al.}{2013}]{hur13}
{Hurley} J.~R.,  {Tout} C.~A.,   {Pols} O.~R.,  2013, {BSE: Binary Star
  Evolution} (\mn@eprint {ascl} {1303.014})

\bibitem[\protect\citeauthoryear{{Ibano{\v{g}}lu}, {Soydugan}, {Soydugan}  \&
  {Dervi{\c{s}}o{\v{g}}lu}}{{Ibano{\v{g}}lu} et~al.}{2006}]{iba06}
{Ibano{\v{g}}lu} C.,  {Soydugan} F.,  {Soydugan} E.,   {Dervi{\c{s}}o{\v{g}}lu}
  A.,  2006, \mn@doi [\mnras] {10.1111/j.1365-2966.2006.11052.x}, \href
  {https://ui.adsabs.harvard.edu/abs/2006MNRAS.373..435I} {373, 435}

\bibitem[\protect\citeauthoryear{{Johnston}}{{Johnston}}{2021}]{joh21}
{Johnston} C.,  2021, \mn@doi [\aap] {10.1051/0004-6361/202141080}, \href
  {https://ui.adsabs.harvard.edu/abs/2021A&A...655A..29J} {655, A29}

\bibitem[\protect\citeauthoryear{Krzywinski \& Altman}{Krzywinski \&
  Altman}{2014}]{krz14}
Krzywinski M.,  Altman N.,  {2014}, \mn@doi [{NATURE METHODS}]
  {{10.1038/nmeth.2813}}, {11}, 119

\bibitem[\protect\citeauthoryear{{Kurucz}}{{Kurucz}}{1979}]{kur79}
{Kurucz} R.~L.,  1979, \mn@doi [\apjs] {10.1086/190589}, \href
  {https://ui.adsabs.harvard.edu/abs/1979ApJS...40....1K} {40, 1}

\bibitem[\protect\citeauthoryear{{Lampens}}{{Lampens}}{2021}]{lam21}
{Lampens} P.,  2021, \mn@doi [Galaxies] {10.3390/galaxies9020028}, \href
  {https://ui.adsabs.harvard.edu/abs/2021Galax...9...28L} {9, 28}

\bibitem[\protect\citeauthoryear{{Lampens}, {Mkrtichian}, {Lehmann},
  {Gunsriwiwat}, {Vermeylen}, {Matthews}  \& {Kuschnig}}{{Lampens}
  et~al.}{2022}]{lam22}
{Lampens} P.,  {Mkrtichian} D.,  {Lehmann} H.,  {Gunsriwiwat} K.,  {Vermeylen}
  L.,  {Matthews} J.,   {Kuschnig} R.,  2022, \mn@doi [\mnras]
  {10.1093/mnras/stac289}, \href
  {https://ui.adsabs.harvard.edu/abs/2022MNRAS.512..917L} {512, 917}

\bibitem[\protect\citeauthoryear{{Lenz} \& {Breger}}{{Lenz} \&
  {Breger}}{2005}]{len05}
{Lenz} P.,  {Breger} M.,  2005, \mn@doi [Communications in Asteroseismology]
  {10.1553/cia146s53}, \href
  {https://ui.adsabs.harvard.edu/abs/2005CoAst.146...53L} {146, 53}

\bibitem[\protect\citeauthoryear{{Liakos}}{{Liakos}}{2015}]{lia15a}
{Liakos} A.,  2015, in {Rucinski} S.~M.,  {Torres} G.,   {Zejda} M.,  eds,
  Astronomical Society of the Pacific Conference Series Vol. 496, Living
  Together: Planets, Host Stars and Binaries. p.~286

\bibitem[\protect\citeauthoryear{{Liakos} \& {Niarchos}}{{Liakos} \&
  {Niarchos}}{2017}]{lia17}
{Liakos} A.,  {Niarchos} P.,  2017, \mn@doi [\mnras] {10.1093/mnras/stw2756},
  \href {https://ui.adsabs.harvard.edu/abs/2017MNRAS.465.1181L} {465, 1181}

\bibitem[\protect\citeauthoryear{{Lucy}}{{Lucy}}{1967}]{luc67}
{Lucy} L.~B.,  1967, \zap, \href
  {https://ui.adsabs.harvard.edu/abs/1967ZA.....65...89L} {65, 89}

\bibitem[\protect\citeauthoryear{{Mkrtichian}, {Kusakin}, {Gamarova}  \&
  {Nazarenko}}{{Mkrtichian} et~al.}{2002}]{mkr02}
{Mkrtichian} D.~E.,  {Kusakin} A.~V.,  {Gamarova} A.~Y.,   {Nazarenko} V.,
  2002, in {Aerts} C.,  {Bedding} T.~R.,   {Christensen-Dalsgaard} J.,  eds,
  Astronomical Society of the Pacific Conference Series Vol. 259, IAU Colloq.
  185: Radial and Nonradial Pulsationsn as Probes of Stellar Physics. p.~96

\bibitem[\protect\citeauthoryear{{Mortensen}, {Eisner}, {IJspeert}, {Kochoska}
  \& {Prsa}}{{Mortensen} et~al.}{2021}]{mor21}
{Mortensen} D.,  {Eisner} N.,  {IJspeert} L.,  {Kochoska} A.,   {Prsa} A.,
  2021, in American Astronomical Society Meeting Abstracts. p. 530.01

\bibitem[\protect\citeauthoryear{{Ochsenbein}, {Bauer}  \&
  {Marcout}}{{Ochsenbein} et~al.}{2000}]{och00}
{Ochsenbein} F.,  {Bauer} P.,   {Marcout} J.,  2000, \mn@doi [\aaps]
  {10.1051/aas:2000169}, \href
  {https://ui.adsabs.harvard.edu/abs/2000A&AS..143...23O} {143, 23}

\bibitem[\protect\citeauthoryear{{Okazaki}}{{Okazaki}}{1977}]{oka77}
{Okazaki} A.,  1977, \pasj, \href
  {https://ui.adsabs.harvard.edu/abs/1977PASJ...29..289O} {29, 289}

\bibitem[\protect\citeauthoryear{{{\"O}zdarcan} \&
  {{\c{C}}akirli}}{{{\"O}zdarcan} \& {{\c{C}}akirli}}{2020}]{ozd20}
{{\"O}zdarcan} O.,  {{\c{C}}akirli} {\"O}.,  2020, \mn@doi [\rmxaa]
  {10.22201/ia.01851101p.2020.56.02.12}, \href
  {https://ui.adsabs.harvard.edu/abs/2020RMxAA..56..321O} {56, 321}

\bibitem[\protect\citeauthoryear{{Peters} \& {Wilson}}{{Peters} \&
  {Wilson}}{2012}]{pet12}
{Peters} G.~J.,  {Wilson} R.~E.,  2012, in American Astronomical Society
  Meeting Abstracts \#220. p. 406.04

\bibitem[\protect\citeauthoryear{{Pr{\v{s}}a} \& {Zwitter}}{{Pr{\v{s}}a} \&
  {Zwitter}}{2005}]{pri05}
{Pr{\v{s}}a} A.,  {Zwitter} T.,  2005, \mn@doi [\apj] {10.1086/430591}, \href
  {https://ui.adsabs.harvard.edu/abs/2005ApJ...628..426P} {628, 426}

\bibitem[\protect\citeauthoryear{{Reiners} \& {Zechmeister}}{{Reiners} \&
  {Zechmeister}}{2020}]{rei20}
{Reiners} A.,  {Zechmeister} M.,  2020, \mn@doi [\apjs]
  {10.3847/1538-4365/ab609f}, \href
  {https://ui.adsabs.harvard.edu/abs/2020ApJS..247...11R} {247, 11}

\bibitem[\protect\citeauthoryear{{Ribas}, {Jordi}  \& {Gim{\'e}nez}}{{Ribas}
  et~al.}{2000}]{rib00}
{Ribas} I.,  {Jordi} C.,   {Gim{\'e}nez} {\'A}.,  2000, \mn@doi [\mnras]
  {10.1046/j.1365-8711.2000.04035.x}, \href
  {https://ui.adsabs.harvard.edu/abs/2000MNRAS.318L..55R} {318, L55}

\bibitem[\protect\citeauthoryear{{Ricker} et~al.,}{{Ricker}
  et~al.}{2015}]{ric15}
{Ricker} G.~R.,  et~al., 2015, \mn@doi [Journal of Astronomical Telescopes,
  Instruments, and Systems] {10.1117/1.JATIS.1.1.014003}, \href
  {https://ui.adsabs.harvard.edu/abs/2015JATIS...1a4003R} {1, 014003}

\bibitem[\protect\citeauthoryear{{Rodr{\'\i}guez} et~al.,}{{Rodr{\'\i}guez}
  et~al.}{2004}]{rod04}
{Rodr{\'\i}guez} E.,  et~al., 2004, \mn@doi [\mnras]
  {10.1111/j.1365-2966.2004.07314.x}, \href
  {https://ui.adsabs.harvard.edu/abs/2004MNRAS.347.1317R} {347, 1317}

\bibitem[\protect\citeauthoryear{{Ruci{\'n}ski}}{{Ruci{\'n}ski}}{1969}]{ruc69}
{Ruci{\'n}ski} S.~M.,  1969, \actaa, \href
  {https://ui.adsabs.harvard.edu/abs/1969AcA....19..245R} {19, 245}

\bibitem[\protect\citeauthoryear{{Samadi Ghadim}, {Lampens}  \&
  {Jassur}}{{Samadi Ghadim} et~al.}{2018}]{sam18}
{Samadi Ghadim} A.,  {Lampens} P.,   {Jassur} M.,  2018, \mn@doi [\mnras]
  {10.1093/mnras/stx3072}, \href
  {https://ui.adsabs.harvard.edu/abs/2018MNRAS.474.5549S} {474, 5549}

\bibitem[\protect\citeauthoryear{{Sharma} et~al.,}{{Sharma}
  et~al.}{2018}]{sha18}
{Sharma} S.,  et~al., 2018, \mn@doi [\mnras] {10.1093/mnras/stx2582}, \href
  {https://ui.adsabs.harvard.edu/abs/2018MNRAS.473.2004S} {473, 2004}

\bibitem[\protect\citeauthoryear{{Shaw}}{{Shaw}}{1994}]{sha94}
{Shaw} J.~S.,  1994, \memsai, \href
  {https://ui.adsabs.harvard.edu/abs/1994MmSAI..65...95S} {65, 95}

\bibitem[\protect\citeauthoryear{{Sim{\'o}n-D{\'\i}az}, {Godart}, {Castro},
  {Herrero}, {Aerts}, {Puls}, {Telting}  \&
  {Grassitelli}}{{Sim{\'o}n-D{\'\i}az} et~al.}{2017}]{sim17}
{Sim{\'o}n-D{\'\i}az} S.,  {Godart} M.,  {Castro} N.,  {Herrero} A.,  {Aerts}
  C.,  {Puls} J.,  {Telting} J.,   {Grassitelli} L.,  2017, \mn@doi [\aap]
  {10.1051/0004-6361/201628541}, \href
  {https://ui.adsabs.harvard.edu/abs/2017A&A...597A..22S} {597, A22}

\bibitem[\protect\citeauthoryear{{Southworth} \& {Bowman}}{{Southworth} \&
  {Bowman}}{2022}]{sou22}
{Southworth} J.,  {Bowman} D.~M.,  2022, \mn@doi [\mnras]
  {10.1093/mnras/stac875}, \href
  {https://ui.adsabs.harvard.edu/abs/2022MNRAS.513.3191S} {513, 3191}

\bibitem[\protect\citeauthoryear{{Soydugan}, {Frasca}, {Soydugan}, {Catalano},
  {Demircan}  \& {Ibano{\v{g}}lu}}{{Soydugan} et~al.}{2007}]{soy07}
{Soydugan} F.,  {Frasca} A.,  {Soydugan} E.,  {Catalano} S.,  {Demircan} O.,
  {Ibano{\v{g}}lu} C.,  2007, \mn@doi [\mnras]
  {10.1111/j.1365-2966.2007.12065.x}, \href
  {https://ui.adsabs.harvard.edu/abs/2007MNRAS.379.1533S} {379, 1533}

\bibitem[\protect\citeauthoryear{{Stassun} et~al.,}{{Stassun}
  et~al.}{2019}]{sta19}
{Stassun} K.~G.,  et~al., 2019, \mn@doi [\aj] {10.3847/1538-3881/ab3467}, \href
  {https://ui.adsabs.harvard.edu/abs/2019AJ....158..138S} {158, 138}

\bibitem[\protect\citeauthoryear{{Terrell} \& {Wilson}}{{Terrell} \&
  {Wilson}}{2005}]{ter05}
{Terrell} D.,  {Wilson} R.~E.,  2005, \mn@doi [\apss]
  {10.1007/s10509-005-4449-4}, \href
  {https://ui.adsabs.harvard.edu/abs/2005Ap&SS.296..221T} {296, 221}

\bibitem[\protect\citeauthoryear{{Tkachenko} et~al.,}{{Tkachenko}
  et~al.}{2020}]{tka20}
{Tkachenko} A.,  et~al., 2020, \mn@doi [\aap] {10.1051/0004-6361/202037452},
  \href {https://ui.adsabs.harvard.edu/abs/2020A&A...637A..60T} {637, A60}

\bibitem[\protect\citeauthoryear{{Torres}, {Andersen}  \&
  {Gim{\'e}nez}}{{Torres} et~al.}{2010}]{tor10}
{Torres} G.,  {Andersen} J.,   {Gim{\'e}nez} A.,  2010, \mn@doi [\aapr]
  {10.1007/s00159-009-0025-1}, \href
  {https://ui.adsabs.harvard.edu/abs/2010A&ARv..18...67T} {18, 67}

\bibitem[\protect\citeauthoryear{{Ula{\c{s}}}, {Ulusoy}, {Erkan}, {Madiba}  \&
  {Matsete}}{{Ula{\c{s}}} et~al.}{2022}]{ula22}
{Ula{\c{s}}} B.,  {Ulusoy} C.,  {Erkan} N.,  {Madiba} M.,   {Matsete} M.,
  2022, \mn@doi [\apss] {10.1007/s10509-022-04053-6}, \href
  {https://ui.adsabs.harvard.edu/abs/2022Ap&SS.367...22U} {367, 22}

\bibitem[\protect\citeauthoryear{{Uytterhoeven} et~al.,}{{Uytterhoeven}
  et~al.}{2011}]{uyt11}
{Uytterhoeven} K.,  et~al., 2011, \mn@doi [\aap] {10.1051/0004-6361/201117368},
  \href {https://ui.adsabs.harvard.edu/abs/2011A&A...534A.125U} {534, A125}

\bibitem[\protect\citeauthoryear{{Virnina}, {Andronov}  \&
  {Mogorean}}{{Virnina} et~al.}{2011}]{vir11}
{Virnina} N.~A.,  {Andronov} I.~L.,   {Mogorean} M.~V.,  2011, Journal of
  Physical Studies, \href
  {https://ui.adsabs.harvard.edu/abs/2011JPhSt..15.2901V} {15, 2901}

\bibitem[\protect\citeauthoryear{{Von Zeipel}}{{Von Zeipel}}{1924}]{zei24}
{Von Zeipel} H.,  1924, \mn@doi [\mnras] {10.1093/mnras/84.9.665}, \href
  {https://ui.adsabs.harvard.edu/abs/1924MNRAS..84..665V} {84, 665}

\bibitem[\protect\citeauthoryear{{Walter}}{{Walter}}{1970}]{wal70}
{Walter} K.,  1970, \aap, \href
  {https://ui.adsabs.harvard.edu/abs/1970A&A.....5..140W} {5, 140}

\bibitem[\protect\citeauthoryear{{Walter}}{{Walter}}{1978}]{wal78}
{Walter} K.,  1978, \aaps, \href
  {https://ui.adsabs.harvard.edu/abs/1978A&AS...32...57W} {32, 57}

\bibitem[\protect\citeauthoryear{{Wilson} \& {Devinney}}{{Wilson} \&
  {Devinney}}{1971}]{wil71}
{Wilson} R.~E.,  {Devinney} E.~J.,  1971, \mn@doi [\apj] {10.1086/150986},
  \href {https://ui.adsabs.harvard.edu/abs/1971ApJ...166..605W} {166, 605}

\bibitem[\protect\citeauthoryear{{Wilson}, {Devinney}  \& {Van Hamme}}{{Wilson}
  et~al.}{2020}]{wil20}
{Wilson} R.~E.,  {Devinney} E.~J.,   {Van Hamme} W.,  2020, {WD:
  Wilson-Devinney binary star modeling} (\mn@eprint {ascl} {2004.004})

\bibitem[\protect\citeauthoryear{{Wright}, {Egan}, {Kraemer}  \&
  {Price}}{{Wright} et~al.}{2003}]{wri03}
{Wright} C.~O.,  {Egan} M.~P.,  {Kraemer} K.~E.,   {Price} S.~D.,  2003,
  \mn@doi [\aj] {10.1086/345511}, \href
  {https://ui.adsabs.harvard.edu/abs/2003AJ....125..359W} {125, 359}

\bibitem[\protect\citeauthoryear{{Yang}, {Li}  \& {Dai}}{{Yang}
  et~al.}{2012}]{yan12}
{Yang} Y.~G.,  {Li} L.~H.,   {Dai} H.~F.,  2012, \mn@doi [\aj]
  {10.1088/0004-6256/144/2/50}, \href
  {https://ui.adsabs.harvard.edu/abs/2012AJ....144...50Y} {144, 50}

\makeatother
\end{thebibliography}



\appendix

\section{List of combined frequencies}

\setcounter{table}{0}
\renewcommand\thetable{\Alph{section}.\arabic{table}}

Table~\ref{ta1} lists the possible combination frequencies as a continuation of Table~\ref{gentab}. The combination relations are denoted in the first column.

\onecolumn

\begin{center}
\small
\begin{longtable}{lcccc}
\caption{Combination frequencies derived from the frequency analysis. The standard errors are given in parentheses for the last digits.} \label{ta1}
\\

\hline 
& $f$ (d$^{-1}$) & $A$ (mmag) & $\phi$  &  SNR \\

\hline 
\endfirsthead

\multicolumn{3}{c}%
{{\tablename\ \thetable{} -- continued from previous page}} \\
\hline 
& $f$ (d$^{-1}$) & $A$ (mmag) & $\phi$  &  SNR \\ \hline 
\endhead

\hline \multicolumn{4}{r}{{Continued on next page}} \\
\endfoot

\hline 
\endlastfoot
IQ~CMa &  & &   &   \\
\cmidrule{1-1}
$f_{1} \approx 4f_{orb}$&5.46937(7)&0.00570(2)&0.5759(5)&142.7\\
$f_{2} \approx 6f_{orb}$&8.20212(11)&0.00343(2)&0.0246(8)&101.7\\
$f_{3} \approx f_{orb}$&1.36541(26)&0.00144(2)&0.4804(19)&23.0\\
$f_{4} \approx 5f_{orb}$&6.83865(26)&0.00141(2)&0.2889(20)&43.3\\
$f_{6} \approx f_{1}+f_{3}-f_{4}$&0.02711(51)&0.00072(2)&0.8840(38)&12.3\\
$f_{7} \approx f_{4}+2f_{2}$&23.26032(51)&0.00073(2)&0.7394(38)&14.6\\
$f_{9} \approx 8f_{orb}$&10.93681(31)&0.00119(2)&0.1912(23)&35.8\\
$f_{11} \approx f_{4}+2f_{10}-2f_{5}$&33.46698(47)&0.00080(2)&0.1437(35)&15.9\\
$f_{12} \approx f_{4}+2f_{7}-f_{5}$&35.44246(75)&0.00050(2)&0.9330(56)&12.0\\
$f_{13} \approx f_{1}+2f_{7}-f_{11}$&18.52692(75)&0.00050(2)&0.6818(56)&10.7\\
$f_{14} \approx f_{1}+f_{11}-f_{2}$&30.73229(68)&0.00054(2)&0.0267(51)&12.1\\
$f_{15} \approx f_{1}+f_{14}$&36.15905(90)&0.00041(2)&0.2796(67)&9.7\\
$f_{16} \approx f_{2}+f_{3}$&9.57915(69)&0.00054(2)&0.5547(51)&16.1\\
$f_{17} \approx f_{1}+f_{4}+f_{5}$&30.21905(120)&0.00031(2)&0.4783(89)&6.8\\
$f_{18} \approx f_{17}-f_{2}$&21.99756(121)&0.00031(2)&0.3646(90)&4.9\\
$f_{19} \approx f_{14}+f_{5}-f_{1}$&43.11197(130)&0.00029(2)&0.2495(97)&6.5\\
$f_{20} \approx f_{1}$&5.49842(115)&0.00032(2)&0.7803(86)&8.1\\
$f_{21} \approx f_{10}+f_{2}-f_{6}$&39.37599(153)&0.00024(2)&0.5282(114)&5.1\\
$f_{22} \approx f_{13}+f_{4}$&25.38881(132)&0.00028(2)&0.1235(98)&5.0\\
$f_{23} \approx f_{1}+f_{12}-f_{8}$&19.56888(141)&0.00026(2)&0.0711(105)&4.0\\
$f_{24} \approx f_{1}+f_{22}$&30.87755(152)&0.00024(2)&0.6541(113)&5.3\\
$f_{25} \approx f_{16}+f_{17}$&39.82531(161)&0.00023(2)&0.9758(120)&4.8\\
$f_{26} \approx 2f_{16}$&19.16217(121)&0.00031(2)&0.9508(90)&5.4\\
$f_{27} \approx 12f_{orb}$&16.40618(135)&0.00028(2)&0.0594(101)&6.9\\
$f_{28} \approx f_{21}-f_{2}$&31.16806(154)&0.00024(2)&0.6410(115)&5.1\\
$f_{29} \approx f_{19}-f_{1}$&37.66778(163)&0.00023(2)&0.9586(122)&5.3\\
\hline
AW~Men &  & &   &   \\
\cmidrule{1-1}
$f_{3} \approx 4f_{1}$&0.03569(2)&0.00759(3)&0.8176(6)&52.5\\
$f_{4} \approx f_{orb}$&0.20615(2)&0.00681(3)&0.4094(7)&48.5\\
$f_{5} \approx f_{3}+2f_{4}$&0.43671(3)&0.00494(3)&0.5827(10)&35.6\\
$f_{6} \approx 2f_{3}$&0.07185(7)&0.00223(3)&0.0062(21)&15.6\\
$f_{7} \approx f_{2}-f_{3}-f_{4}$&11.31597(5)&0.00321(3)&0.8109(15)&42.6\\
$f_{8} \approx f_{1}+f_{6}$&0.08922(3)&0.00539(3)&0.6370(9)&37.6\\
$f_{9} \approx 2f_{6}$&0.15778(2)&0.00658(3)&0.2456(7)&46.4\\
$f_{10} \approx 3f_{3}$&0.11458(4)&0.00439(3)&0.4537(11)&30.6\\
$f_{11} \approx f_{1}+f_{3}$&0.05353(4)&0.00407(3)&0.9726(12)&28.1\\
$f_{12} \approx 2f_{10}$&0.24043(5)&0.00297(3)&0.5389(16)&21.3\\
$f_{13} \approx f_{9}$&0.14557(3)&0.00505(3)&0.4364(9)&35.5\\
$f_{14} \approx f_{6}$&0.06198(5)&0.00308(3)&0.2726(15)&21.4\\
$f_{15} \approx f_{10}$&0.12350(10)&0.00150(3)&0.9891(32)&10.5\\
$f_{16} \approx f_{5}-f_{14}$&0.38412(9)&0.00176(3)&0.7051(27)&12.8\\
$f_{17} \approx f_{2}+2f_{16}+2f_{5}$&13.22529(12)&0.00127(3)&0.9510(38)&22.2\\
$f_{18} \approx 2f_{8}$&0.18220(9)&0.00165(3)&0.4066(29)&11.8\\
$f_{19} \approx 2f_{13}$&0.28738(5)&0.00286(3)&0.0143(17)&20.5\\
$f_{20} \approx f_{17}-f_{10}-f_{2}$&1.53835(15)&0.00107(3)&0.6135(45)&7.2\\
$f_{21} \approx f_{20}-f_{5}$&1.09835(13)&0.00119(3)&0.4384(40)&7.9\\
$f_{22} \approx 2f_{12}$&0.47240(17)&0.00094(3)&0.9167(51)&6.8\\
$f_{23} \approx f_{12}$&0.25170(6)&0.00250(3)&0.7829(19)&18.0\\
$f_{24} \approx f_{19}$&0.27564(6)&0.00258(3)&0.5498(19)&18.5\\
$f_{25} \approx f_{13}+f_{4}$&0.34373(8)&0.00187(3)&0.8126(26)&13.7\\
$f_{26} \approx f_{25}$&0.33481(10)&0.00155(3)&0.4440(31)&11.3\\
$f_{27} \approx 2f_{7}-f_{4}-2f_{20}$&19.36037(21)&0.00073(3)&0.9400(66)&16.7\\
$f_{28} \approx f_{5}+f_{9}$&0.60106(14)&0.00112(3)&0.9976(43)&7.9\\
$f_{29} \approx f_{27}+2f_{21}$&21.56835(23)&0.00068(3)&0.0007(70)&14.7\\
$f_{30} \approx f_{10}+f_{21}+f_{27}$&20.57988(23)&0.00068(3)&0.8801(70)&16.2\\
$f_{31} \approx f_{5}+f_{8}$&0.53204(18)&0.00088(3)&0.2344(54)&6.3\\
$f_{32} \approx f_{29}-f_{2}$&9.99269(25)&0.00063(3)&0.4504(76)&10.0\\
$f_{33} \approx 2f_{26}$&0.67291(18)&0.00085(3)&0.8258(56)&5.9\\
$f_{34} \approx f_{19}+f_{5}$&0.73255(19)&0.00083(3)&0.1986(58)&5.7\\
$f_{35} \approx f_{25}$&0.35313(14)&0.00113(3)&0.6385(42)&8.2\\
$f_{36} \approx 2f_{4}$&0.40666(20)&0.00079(3)&0.4207(60)&5.8\\
$f_{37} \approx f_{4}$&0.21319(17)&0.00091(3)&0.4889(52)&6.5\\
$f_{38} \approx f_{8}$&0.08124(11)&0.00139(3)&0.9754(34)&9.7\\
$f_{39} \approx f_{30}-f_{8}$&20.49535(31)&0.00051(3)&0.7385(94)&12.2\\
$f_{40} \approx f_{20}+f_{5}$&1.97647(27)&0.00057(3)&0.3867(84)&4.3\\
$f_{41} \approx f_{2}+f_{35}$&11.93347(31)&0.00050(3)&0.6050(96)&6.9\\
\hline
W~Vol &  & &   &   \\
\cmidrule{1-1}
$f_{1} \approx 4f_{orb}$&1.45015(1)&0.00717(2)&0.198(0)&113.2\\
$f_{4} \approx 6f_{orb}$&2.17540(2)&0.00463(2)&0.926(1)&85.3\\
$f_{5} \approx 5f_{orb}$&1.81315(6)&0.00120(2)&0.579(2)&20.8\\
$f_{6} \approx 2f_{2}$&0.02164(5)&0.00131(2)&0.423(2)&17.0\\
$f_{7} \approx f_{orb}$&0.35964(4)&0.00168(2)&0.647(2)&21.9\\
$f_{8} \approx 8f_{orb}$&2.90066(4)&0.00168(2)&0.641(2)&37.5\\
$f_{9} \approx f_{orb}$&0.36524(5)&0.00135(2)&0.018(2)&17.6\\
$f_{12} \approx 4f_{6}$&0.08432(9)&0.00075(2)&0.437(4)&9.8\\
$f_{15} \approx f_{3}+f_{7}-f_{6}$&20.08453(14)&0.00049(2)&0.050(5)&6.2\\
$f_{16} \approx f_{13}+2f_{12}+2f_{8}$&21.32090(16)&0.00044(2)&0.579(6)&6.5\\
$f_{17} \approx f_{12}+f_{16}-f_{5}$&19.59729(13)&0.00052(2)&0.744(5)&6.6\\
$f_{18} \approx f_{12}+f_{3}$&19.46186(15)&0.00047(2)&0.050(6)&6.1\\
$f_{19} \approx f_{15}-f_{8}$&17.19133(19)&0.00037(2)&0.422(7)&5.8\\
$f_{20} \approx f_{13}+f_{7}$&15.71805(15)&0.00046(2)&0.003(6)&7.1\\
$f_{21} \approx f_{1}+f_{5}$&3.26255(18)&0.00038(2)&0.173(7)&8.8\\
$f_{22} \approx 2f_{4}$&4.35081(19)&0.00036(2)&0.101(7)&9.3\\
$f_{23} \approx f_{19}-f_{7}$&16.82832(20)&0.00035(2)&0.571(8)&5.3\\
$f_{24} \approx f_{16}-f_{1}-f_{12}$&19.78943(21)&0.00033(2)&0.891(8)&4.3\\
$f_{25} \approx f_{16}+f_{24}-f_{11}$&20.71241(22)&0.00032(2)&0.059(8)&4.1\\
$f_{26} \approx f_{12}-f_{6}$&0.05745(17)&0.00041(2)&0.920(7)&5.5\\
$f_{27} \approx f_{3}+f_{7}-f_{17}$&0.15706(21)&0.00033(2)&0.883(8)&4.3\\
$f_{28} \approx f_{7}-f_{26}$&0.30219(24)&0.00029(2)&0.392(9)&4.0\\
$f_{29} \approx f_{10}+f_{12}+f_{4}$&18.56835(22)&0.00031(2)&0.066(9)&4.5\\
$f_{30} \approx f_{19}-f_{1}$&15.74230(21)&0.00032(2)&0.313(8)&5.1\\
$f_{31} \approx f_{16}-f_{2}$&21.30374(22)&0.00031(2)&0.978(9)&4.6\\
$f_{32} \approx 2f_{orb}$&0.71146(18)&0.00038(2)&0.194(7)&5.2\\
$f_{33} \approx f_{14}+f_{8}$&18.73138(23)&0.00030(2)&0.942(9)&4.3\\
$f_{34} \approx f_{31}-f_{27}$&21.13548(20)&0.00035(2)&0.493(8)&5.1\\
$f_{35} \approx f_{10}-f_{28}$&16.00718(26)&0.00027(2)&0.220(10)&4.2\\
\hline
\end{longtable}
\end{center}

\twocolumn




\bsp	
\label{lastpage}
\end{document}